\documentclass[preprint,journal,twoside]{IEEEtran}
\usepackage[table]{xcolor}
\usepackage{cite}
\usepackage{dsfont}
\usepackage{amsmath,amssymb,amsfonts}
\usepackage{mathtools}
\usepackage{bm}
\usepackage{float}
\usepackage{caption}
\usepackage{subcaption}
\usepackage{setspace}

\usepackage{amsthm}
\usepackage[ruled,linesnumbered]{algorithm2e}
\usepackage{bigints}
\usepackage{relsize}
\usepackage{braket}
\usepackage{graphicx}
\usepackage{textcomp}
\usepackage{stmaryrd}
\usepackage{array,booktabs,,dcolumn,caption}
\usepackage{tabularx}
\usepackage{multirow}
\usepackage[colorlinks=true, allcolors=blue]{hyperref}
\usepackage{setspace}
\def\BState{\State\hskip-\ALG@thistlm}
\theoremstyle{definition}

\newcolumntype{d}[1]{D{.}{.}{#1}} % "decimal" column type
\makeatletter
\newcolumntype{B}[3]{>{\boldmath\DC@{#1}{#2}{#3}}c<{\DC@end}}
\renewcommand{\ast}{{}^{\textstyle *}} % for raised "asterisks"
\renewcommand{\vec}[1]{\mathbf{#1}}

\newcommand{\coeff}{\vec{f}}
\newcommand{\cmeas}{\coeff_{meas}}
\newcommand{\hcmeas}{\hat{\coeff}_{meas}}
\newcommand{\cnull}{\coeff_{null}}
\newcommand{\hcnull}{\hat{\coeff}_{null}}
\newcommand{\hcoeff}{\hat{\vec{f}}}

\newcommand{\proj}{\mathcal{P}_{\mathcal{A}}(\vec{v})}
\newcommand{\cUM}{\hcoeff^{UM}}
\newcommand{\cUMmeas}{\hat{\coeff}^{UM}_{meas}}
\newcommand{\cUMnull}{\hat{\coeff}^{UM}_{null}}

\newcommand{\clearsubcaptcounter}{\setcounter{sub\@captype}{0}}
\newtheorem{theorem}{Theorem}[section]

\DeclareMathOperator*{\argmin}{argmin}

\DeclareMathAlphabet\mathbfcal{OMS}{cmsy}{b}{n}

\begin{document}

\title{Mining the manifolds of deep generative models for multiple data-consistent solutions of ill-posed tomographic imaging problems}
%\title{Mining the manifold of deep generative models for diverse images from tomographic measurements}
\author{Sayantan Bhadra,
        Umberto Villa,~\IEEEmembership{Member, IEEE},
        and Mark A. Anastasio,~\IEEEmembership{Senior Member, IEEE}% <-this % stops a space
\thanks{This work was supported in part by NIH Awards EB020604, EB023045, NS102213, EB028652, and NSF Award DMS1614305.}
\thanks{Sayantan Bhadra is 
with the Department of Computer Science and Engineering, Washington University in St. Louis, St. Louis, 
MO 63130 USA (e-mail: sayantanbhadra@wustl.edu).}
\thanks{Umberto Villa is with 
the Oden Institute, The University of Texas at Austin, Austin, TX 78712 USA (e-mail: uvilla@austin.utexas.edu).}
\thanks{Mark A. Anastasio is with 
the Department of Bioengineering, University of Illinois at Urbana-Champaign, Urbana, IL 61801 USA (e-mail: maa@illinois.edu).}}

% \markboth{Journal of \LaTeX\ Class Files,~Vol.~14, No.~8, August~2015}{Bhadra \MakeLowercase{\textit{et al.}}: Mining the manifold of deep generative models}%
% %{Shell \MakeLowercase{\textit{et al.}}: Bare Demo of IEEEtran.cls for IEEE Journals}

% make the title area
\maketitle

% As a general rule, do not put math, special symbols or citations
% in the abstract or keywords.
\begin{abstract}
Tomographic imaging is in general an ill-posed inverse problem. Typically, a single regularized image estimate of the sought-after object is obtained from tomographic measurements. However, there may be multiple objects that are all consistent with the same measurement data. The ability to generate such alternate solutions is important because it may enable new assessments of imaging systems. In principle, this can be achieved by means of posterior sampling methods. In recent years, deep neural networks have been employed for posterior sampling with promising results. However, such methods are not yet for use with large-scale tomographic imaging applications. On the other hand, empirical sampling methods may be computationally feasible for large-scale imaging systems and enable uncertainty quantification for practical applications. Empirical sampling involves solving a regularized inverse problem within a stochastic optimization framework to obtain alternate data-consistent solutions. In this work, a new empirical sampling method is proposed that computes multiple solutions of a tomographic inverse problem that are consistent with the same acquired measurement data. The method operates by repeatedly solving an optimization problem in the latent space of a style-based generative adversarial network (StyleGAN), and was inspired by the Photo Upsampling via Latent Space Exploration (PULSE) method that was developed for super-resolution tasks. The proposed method is demonstrated and analyzed via numerical studies that involve two stylized tomographic imaging modalities. These studies establish the ability of the method to perform efficient empirical sampling and uncertainty quantification.
\end{abstract}

% Note that keywords are not normally used for peerreview papers.

\IEEEpeerreviewmaketitle

\section{Introduction}
\label{sec:introduction}

%\IEEEPARstart{A} 
Modern imaging systems are computed in nature and require an appropriate image reconstruction method for estimating an object from a collection of tomographic measurements \cite{kak2002principles}. In practice, the acquired measurement data is noisy, and at times incomplete, in which case the associated inverse problem will be ill-posed. For example, to accelerate the data-acquisition in magnetic resonance imaging, undersampled measurement data can be purposely acquired \cite{zbontar2018fastmri}.  In such cases, image reconstruction methods that seek to estimate approximate but potentially useful estimates of the object property require regularization.  Regularization strategies incorporate appropriate prior knowledge of the object, known as object priors in the Bayesian parlance, into the reconstruction process.  For example, sparsity-promoting regularization strategies have found great success in recent years \cite{candes2006stable, sidky2008image, ravishankar2019image}. More recently, a variety of data-driven methods have been proposed whereby a more accurate object prior is learned from large databases of existing imaging data. Many data-driven methods employ deep neural networks, otherwise known as deep learning (DL) \cite{kelly2017deep,mccann2017convolutional,ravishankar2019image}.

Most medical image reconstruction methods available today are designed to produce a single estimate of the object, which is known as the \emph{maximum a posteriori} (MAP) point estimate when interpreted in a statistical framework. However, in the presence of data noise or incompleteness, multiple objects can exist that are consistent with a given set of measurement data. Moreover, there is generally no guarantee that the produced object estimate will be the most accurate or useful (with respect to a specific clinical task) among the multiple possible objects that are consistent with the measured data. This is especially true for many DL-based image reconstruction methods, which are often based on heuristic designs and can have an enhanced propensity for producing hallucinated structures \cite{bhadra2021hallucinations}. These hallucinated structures are of particular concern for medical imaging applications because such structures may not always be readily identifiable as artifacts and therefore the images can appear plausible but are, in fact, incorrect.  

The ability to identify multiple objects that are consistent with a given set of measurement data is of significant importance to the assessment and refinement of data-acquisition designs and image reconstruction procedures. 
For example, from a collection of distinct data-consistent objects,
 uncertainty maps \cite{tick2016image, shaw2021estimating} can be computed.
 Such  maps can be employed to reveal the reliability of a reconstructed image corresponding to a given data-acquisition design, or employed to estimate various figures-of-merit (FOMs) that describe the likelihood of hallucinated structures \cite{bhadra2021hallucinations}. The ability to identify multiple data-consistent objects could also permit analysis of the impact of the null space of a linear imaging operator in new, problem-specific, ways and enable the design of numerical experiments to reveal image reconstruction instabilities \cite{gottschling2020troublesome}.
Moreover, a new capability to generate ensembles of data-consistent objects is needed to advance task-informed adaptive imaging procedures \cite{clarkson2008task,barrett2008adaptive}. 

The generation of multiple solutions to an inverse problem is consistent with the goal of Bayesian inversion methods \cite{ulrych2001bayes}. 
In imaging applications, this can be conceptually achieved by sampling from the posterior distribution that describes the sought-after object conditioned on a set of measurement data \cite{mohebi2006posterior}. This is a holy grail of image reconstruction, but it remains generally impractical in medical imaging applications due to their large scale \cite{green2015bayesian}. In recent years, computational procedures for accomplishing \textit{approximate} posterior sampling in limited-scale problems have been proposed that employ deep neural networks combined with Markov chain Monte Carlo (MCMC) sampling methods or Langevin dynamics \cite{sun2020deep,herrmann2021KAUSTdbi,mosser2018stochastic,jalal2021robust}. While promising, the efficacy of such methods for use with large-scale medical image reconstruction problems remains a topic of investigation. To circumvent the computational challenges of posterior sampling methods, \textit{empirical sampling}\cite{sun2020deep,bardsley2014randomize,akiyama2019first} can be performed to obtain multiple distinct objects that are consistent with a given set of measurement data. In an empirical sampling method, multiple data-consistent solutions are obtained by solving a regularized inverse problem within a stochastic optimization framework\cite{spall2012stochastic}. Empirical sampling methods, while not guaranteeing true posterior sampling, can be computationally feasible for large-scale imaging systems. Moreover, in contrast to MCMC-based posterior sampling methods in which the samples are generated sequentially, alternate solutions obtained via empirical sampling are independent of one another and can be obtained in parallel, thus providing reductions in computation times.

It may be possible that when multiple solutions are sought from a single acquisition of measurement data, the data-consistent objects may contain unrecognizable structures that are irrelevant to the medical imaging application at hand. Such situations may arise when there is no constraint imposed on the objects to ensure that they are relevant to a specified imaging application. One way to achieve empirical sampling that produces application-relevant and data-consistent objects is to constrain the process by use of a deep generative model \cite{goodfellow2016deep} that characterizes the distribution of to-be-imaged objects. For the single image super-resolution problem (SISR), a technique called Photo Upsampling via Latent Space Exploration (PULSE) was proposed in \cite{menon2020pulse} to generate diverse photorealistic high-resolution images from a single low-resolution image. A state-of-the-art deep generative model known as StyleGAN \cite{karras2019style} was employed to characterize the distribution of high resolution images. However, there remains an important need to extend this method for general tomographic inverse problems and quantitatively investigate the data-consistency of the generated samples.

In this work, the following problem is addressed: \textit{Assume a StyleGAN describing a distribution of to-be-imaged objects, a tomographic measurement model, and a single acquisition of incomplete and noisy measurement data are provided.  Find a collection of distinct objects that are consistent with the same acquired measurement data (in a to-be-prescribed sense) and reside in the range of the StyleGAN.} A key motivation for formulating this problem  is to establish an application-relevant empirical sampling method that can be employed in preliminary assessments and refinements of data-acquisition designs and imaging technologies via virtual imaging trials. A method for solving this problem is proposed that is referred to as the PULSE++ method. The PULSE++ method represents the first extension of the PULSE methodology for use with tomographic imaging problems. By utilizing improved  assumptions about the statistics of object embeddings in the latent space of the StyleGAN, the ability of the PULSE++ method to produce diverse, application-relevant, and data-consistent objects is enhanced as compared to the original PULSE method. It should be noted that PULSE++ may be interpreted as an image reconstruction method if a single ``best'' image estimate is chosen from the data-consistent alternate solutions based on an appropriate image quality metric. However, the goal of this study is not to produce a single image estimate, but to find multiple alternate solutions that are all consistent with the measurement data while being relevant to the imaging application. Hence, in this study, PULSE++ is not utilized as an image reconstruction method, but rather as an efficient empirical sampling technique that can be employed to assess and refine a given data acquisition design in new, problem-specific ways, e.g. by enabling computation of reliable uncertainty maps.
Two different stylized tomographic imaging modalities are systematically studied that also involve different measurement noise distributions. Uncertainty maps that quantify the degree of variability of data-consistent alternate solutions are computed.

The remainder of this paper is organized as follows. In Sec. \ref{sec:Background}, salient features of the StyleGAN and empirical sampling using PULSE are introduced. Section \ref{sec:validation} describes statistical tests that were performed with StyleGANs trained on medical images for scrutinizing existing assumptions about the embeddings of objects in the StyleGAN latent space. Section \ref{sec:pulse++} outlines the proposed PULSE++ method for efficient empirical sampling to yield application-relevant and data-consistent objects. Sections \ref{sec:Numerical studies} and \ref{sec:results} describe the numerical studies that were performed to demonstrate the ability of the proposed method to reconstruct diverse objects from the observed measurement data and perform uncertainty quantification at scale. Finally, a discussion and summary of the work is provided in \ref{sec:discussion}.

\section{Background}
\label{sec:Background}
A discrete-to-discrete (DD) model of a (possibly) nonlinear imaging system is employed that can be described as \cite{barrett2013foundations}
\begin{equation}\label{eq:imaging_DD}
\vec{g} = \mathcal{H}(\coeff) + \vec{n}.
\end{equation}
Here, $\vec{g} \in \mathbb{E}^M$ is the observed measurements, $\coeff \in \mathbb{E}^N$ is a finite-dimensional approximation of the to-be-imaged object, $\vec{n} \in \mathbb{E}^M$ is the measurement noise and $\mathcal{H}:\mathbb{E}^N \rightarrow \mathbb{E}^M$ is the imaging operator. It is assumed that the DD imaging model represented by Eq. \eqref{eq:imaging_DD} is a sufficiently accurate approximation of the true mapping from a continuous object function to discrete measurements. In this study, an ``image'' of an object will refer to the representation of the object as values on a two-dimensional grid of pixels. A data fidelity function $\mathcal{J}(\vec{g},\vec{f})$ measures the discrepancy between the output of the imaging operator $\mathcal{H}(\coeff)$ and the measurement data $\vec{g}$ \cite{barrett2013foundations}. From a statistical perspective, the data fidelity term $\mathcal{J}(\vec{g},\vec{f})$ is the negative log-likelihood function and depends on the measurement noise distribution. For example, in the presence of independent and identically distributed (i.i.d.) Gaussian noise $\vec{n} \sim \mathcal{N}(\vec{0},\sigma^2 \vec{I}_N)$, the data fidelity term has the form $\mathcal{J}(\vec{g},\vec{f}) = \frac{1}{2\sigma^2}\|\vec{g}-\mathcal{H}(\vec{f})\|_2^2.$

\subsection{Salient features of the StyleGAN latent space}
\label{sec:embedding}
State-of-the-art deep generative models such as the StyleGAN\cite{karras2019style} hold the potential for characterizing the distribution of finite-dimensional approximations of to-be-imaged objects \cite{zhou2021learning,hong20213d}. Let $G:\mathbb{R}^k \rightarrow \mathbb{R}^N$ denote a parameterized deep generative model with $L$ layers, where $k \ll N$. The generator network $G(\vec{z})$ is trained such that it maps a $k$-dimensional latent vector $\vec{z} \in \mathcal{Z}$ sampled from a known distribution, such as a standard Gaussian distribution, to an image that is representative of the distribution formed by the training images. The generator network $G$ in a StyleGAN is composed of two networks: a mapping network $G_m$ and a synthesis network $G_s$ \cite{karras2019style}. $G_m$ is a fully connected neural network (FCNN)\cite{goodfellow2016deep} that maps the latent vector $\vec{z}$ to an intermediate latent vector $\vec{w} \in \mathbb{R}^k$. Subsequently, the latent vector $\vec{w}$ is replicated $L$ times, and each duplicate latent vector $\vec{w}$ is passed through a learned affine transformation\cite{huang2017arbitrary} that encodes semantic information and input to one of the $L$ layers in the synthesis network $G_s$. Each such vector that is input to $G_s$ controls a specific style or semantic attribute in the generated image. The collection of $L$ copies of the vector $\vec{w} \in \mathbb{R}^k$ is represented as the latent matrix $\vec{W} \in \mathbb{R}^{k \times L}$ and the corresponding intermediate latent space is denoted as $\mathcal{W}$. Additionally, the StyleGAN contains a set of $L$ latent noise vectors $\boldsymbol{\Phi} \equiv \{\boldsymbol{\phi}_l\}_{l=1}^L$ such that $\boldsymbol{\phi}_l \in \mathbb{R}^{p_l}$, where $p_l = 4^{(1+\lceil\frac{l}{2}\rceil)}$\cite{karras2019style}. Each latent noise vector $\boldsymbol{\phi}_l$ serves as an input to layer $l$ in $G_s$. These latent noise vectors are sampled from standard Gaussian distributions and multiplied by learned scaling factors \cite{karras2019style} that enable additional stochastic variability in the fine details of the generated images. 

In addition to a superior performance in image synthesis, the style-specific control that is gained with a StyleGAN generator can be leveraged to perform meaningful semantic transformations of objects in tomographic imaging applications \cite{kelkar2021prior,fetty2020latent,schutte2021using}. To perform such semantic transformations, an embedding for the given object must be obtained first in the latent space of the StyleGAN. Abdal \textit{et al.}\cite{abdal2019image2stylegan} proposed an efficient embedding algorithm that involved solving an optimization problem in an \emph{extended} latent space $\mathcal{W}^+ \equiv \mathbb{R}^{k \times L}\supset \mathcal{W}$. %The latent space $\mathcal{W}^+ $ was defined by relaxing the equality constraint on the $L$ vectors in $\vec{W} \in \mathbb{R}^{k \times L}$ such that $\mathcal{W}^+ $.
Penalty terms such as $GEOCROSS$\cite{menon2020pulse} have been proposed that promote the embedding in the extended latent space $\mathcal{W}^+$ to be close to the latent space $\mathcal{W}$, which in turn encourages the embedded object to be near the range of the generator network $G_s$ with the latent space $\mathcal{W}$\cite{wulff2020improving}. 

Recently, a number of studies posited that the problem of embedding via optimization may be better conditioned by utilizing a modified StyleGAN latent space that possessed a more well-defined structure. 
It was empirically observed that the application of a certain computationally cheap and invertible transformation $\mathcal{T}:\mathbb{R}^{k \times L} \rightarrow \mathbb{R}^{k \times L}$ produced a matrix $\vec{V} = \mathcal{T}(\vec{W})$ whose $k$-dimensional vectors $\{ \vec{v}_i \}_{i=1}^L$ approximately followed the standard Gaussian distribution $\mathcal{N}(\vec{0},\vec{I}_k)$ \cite{menon2020pulse,wulff2020improving,zhu2020improved}.
In particular, $\mathcal{T}$ is the composition of a leaky rectified linear unit (ReLU)\cite{goodfellow2016deep} with an affine whitening transformation\cite{zhu2020improved}.
In what follows the spaces $\mathcal{V} \subset \mathbb{R}^{k \times L}$ and $\mathcal{V}^+ \equiv \mathbb{R}^{k \times L}$ denote the images through $\mathcal{T}$ of the spaces $\mathcal{W}$ and $\mathcal{W}^+$, respectively. Using this transformation, the synthesis network $G_s$ can be equivalently represented in terms of a generator network $\tilde{G}(\vec{V},\boldsymbol{\Phi}) := G_s(\mathcal{T}^{-1}(\vec{V}),\boldsymbol{\Phi})$.
It was reported in \cite{menon2020pulse,wulff2020improving,zhu2020improved} that performing optimization in the latent space $\mathcal{V}^+$ with the generator network $\tilde{G}$ allowed for the use of simple regularizers and led to more accurate embeddings.

\subsection{Empirical sampling with PULSE}
\label{sec:generative}
Menon \textit{et al.}\cite{menon2020pulse} employed such preferable embedding properties of the StyleGAN latent space $\mathcal{V}^+$ in an SISR task, with the goal of mining the manifold of the generator network $\tilde{G}(\vec{V},\boldsymbol{\Phi})$ to discover multiple photo-realistic high-resolution images that are consistent with the same low-resolution image. The proposed framework was termed as Photo Upsampling via Latent Space Exploration (PULSE). The PULSE algorithm belongs to a broader class of generative model-constrained methods for ill-posed inverse problems in imaging, known as compressed sensing using generative models (CSGM)\cite{bora2017compressed,bhadra2020medical,kelkar2021compressible}. The embedding problem within the CSGM framework is highly non-convex. However, gradient-based methods have been observed to find \emph{good} local optima in a computationally feasible manner, thus providing solutions to the inverse problem that are compatible with the measurement data \cite{bora2017compressed}. 

In SISR, the measured data $\vec{g} \in \mathbb{E}^M$ is a low-resolution version of the sought-after image $\coeff \in \mathbb{E}^N$, where $M \leq N$. The imaging operator $\mathcal{H}: \mathbb{E}^N \rightarrow \mathbb{E}^M$ in SISR is a degradation operator that removes the higher spatial frequencies from $\vec{f}$. The CSGM optimization problem in PULSE was formulated to recover a high-resolution estimate $\hcoeff\equiv \tilde{G}(\hat{\vec{V}},\hat{\boldsymbol{\Phi}})$ from the low-resolution image $\vec{g}$, stated as
\begin{align}\label{eq:pulse_original}
    \hat{\vec{V}},\hat{\boldsymbol{\Phi}} &= \argmin_{\vec{V},\boldsymbol{\Phi}} \Big\{\mathcal{J}(\vec{g},\tilde{G}(\vec{V},\boldsymbol{\Phi})) + \lambda_g GEOCROSS(\vec{V})\Big\}, \nonumber\\
    &s.t. \; \vec{v}_i \in S^{k-1}(\sqrt{k}), \boldsymbol{\phi}_i \in S^{p_i-1}(\sqrt{p_i}) \: \forall i \in \{1..L\}
\end{align}
where $\lambda_g>0$ is a regularization hyperparameter and $S^{d-1}(r) \equiv \{\vec{a} \in \mathbb{R}^{d} \, |\, \|\vec{a}\|_2 =r \}$ denotes the spherical surface of radius $r$ in $d$-dimensions. The motivation for constraining the latent style $\{\vec{v}_i\}$ and noise $\{\boldsymbol{\phi}_i\}$ vectors on such surfaces was based on the ``soap bubble effect'' \cite{menon2020pulse} observed for standard multivariate Gaussian vectors in high-dimensional spaces, as discussed in Sec. \ref{sec:validation}. 
The data fidelity term $\mathcal{J}(\vec{g},\tilde{G}(\vec{V},\boldsymbol{\Phi}))$ was chosen to be a suitable $\ell_p$-norm. The penalty term $GEOCROSS(\vec{V})$ is defined as the sum of pairwise geodesic distances among the $L$ latent vectors in $\vec{V}$ on $S^{k-1}(\sqrt{k})$ \cite{menon2020pulse}.
The optimization problem in Eq. \eqref{eq:pulse_original} was approximately solved using projected gradient descent with the Adam optimizer \cite{kingma2014adam}. Multiple runs of the optimization problem were performed, and for each run, $\vec{V}$ and $\boldsymbol{\Phi}$ were randomly initialized by sampling from standard Gaussian distributions. Due to the high degree of non-convexity in Eq. \eqref{eq:pulse_original}, varying the initialization on each run produced solutions corresponding to different local minima of the optimization problem. Thus, after the completion of all the CSGM runs, a diverse set of photo-realistic high-resolution images could be obtained that were significantly different from each other while being qualitatively consistent with the same observed low-resolution image.

Despite results that were promising qualitatively, a quantitative validation of the method was omitted. First, the choice of the data fidelity function $\mathcal{J}(\vec{g},\tilde{G}(\vec{V},\boldsymbol{\Phi}))$ as an $\ell_p$-norm was not justified in terms of the statistics of the noise distribution in the low-resolution image. Second, the tolerance level for data consistency to accept a high-resolution image estimate was chosen in an arbitrary fashion irrespective of the measurement noise distribution. It is therefore difficult to determine the degree to which data consistency is being preserved. Third, while the PULSE method as well as previous studies such as \cite{wulff2020improving,zhu2020improved,kelkar2021prior} employed the prior assumption that the latent vectors $\{\vec{v}_i\}$ follow a standard multivariate Gaussian distribution, no rigorous quantitative evaluation was performed to justify the accuracy of this ansatz. While the lack of such quantitative validation may still be acceptable for a computer vision task where the objective is to obtain diverse photo-realistic face images from a given low-resolution image, quantitative assessment of the method is critical for a proper assessment of tomographic imaging systems.

\section{Statistical validation of the Gaussianized latent space in StyleGAN}
\label{sec:validation}
Restricting the latent space vectors of StyleGAN to lie on the $S^{k-1}(\sqrt{k})$ sphere
may increase the risk of data inconsistency when searching for alternate solutions from the same measurement data with the PULSE method. In this section, a statistical study is described that demonstrates that the underlying assumption of a Gaussian structure in the StyleGAN latent space $\mathcal{V}^+$ in the PULSE method is, in fact, inaccurate. The validation study was performed using two StyleGAN models trained on medical image datasets (MRI-StyleGAN and CT-StyleGAN), as well as the open-sourced StyleGAN model trained on human face images of size $1024 \times 1024$\cite{karras2019style} (Face-StyleGAN) that was employed in the PULSE method for SISR \cite{menon2020pulse}. The training of StyleGANs was performed by adapting an open-sourced TensorFlow-based code \cite{stylegan}. MRI-StyleGAN was trained using 60,000 axial knee images of size $256 \times 256$ pixels extracted from the NYU fastMRI dataset \cite{zbontar2018fastmri}. CT-StyleGAN was trained using 60,000 X-ray CT chest images of size $512 \times 512$ pixels extracted from the NIH DeepLesion dataset \cite{yan2018deeplesion}.
For both the MRI and CT StyleGAN models, each image was normalized to lie in the range [0,1] prior to training. The default training hyperparameters of the open-sourced StyleGAN model were employed. The MRI-StyleGAN model was trained using 2 NVIDIA TITAN X GPUs, while training of the CT-StyleGAN model was performed using 4 NVIDIA V100 GPUs. Both the models were trained for $\sim$1 day. The MRI-StyleGAN and CT-StyleGAN models along with their pre-trained weights are provided in the code repository that accompanies this paper \cite{pulse_pp}. For a quantitative assessment of the trained StyleGAN models, the Fr\'echet Inception Distance (FID) \cite{heusel2017gans} was computed. A lower value of the FID score for a GAN indicates better generative performance. The FID score in each case was computed using 20,000 training images and 20,000 StyleGAN-generated images. The FID scores for the MRI-StyleGAN and the CT-StyleGAN were 9.71 and 38.65 respectively. These values fall within the range of FID scores observed for GANs trained on standard MRI and CT datasets \cite{skandarani2021gans}.  
The FID score for the Face-StyleGAN as reported in \cite{karras2019style} was 4.40.  It should be noted, however, that a definitive metric for the quantitative evaluation of GANs remains an active area of research \cite{deshpande2021method,borji2021pros,skandarani2021gans,kelkar2022assessing}. 

\begin{figure}[!htb]
    \centering
     \includegraphics[width=\linewidth]{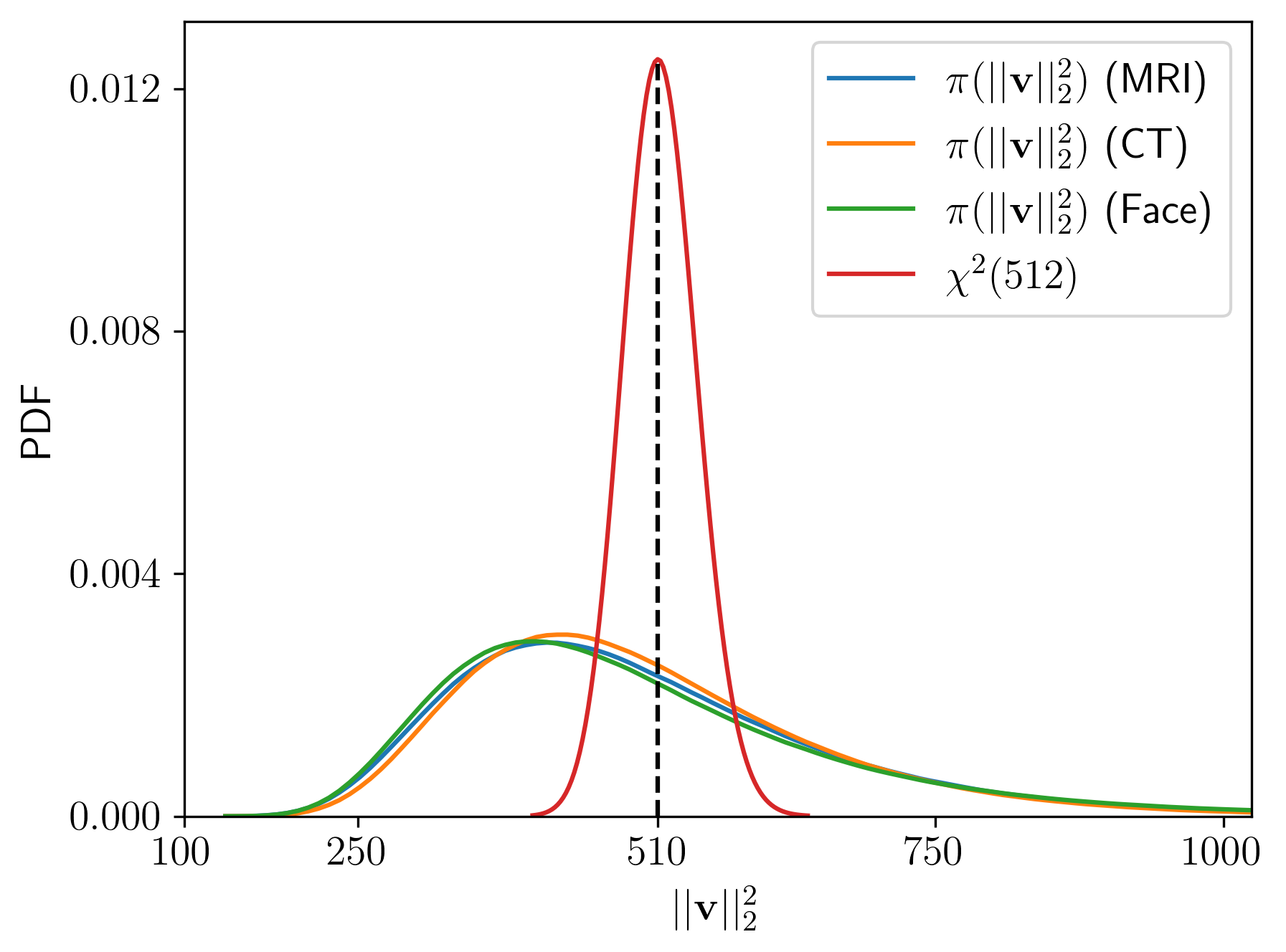}
      \caption{Comparison of  $\pi(\|\vec{v}\|^2_2)$ with the PDF of $\chi^2(k)$ for the MRI, CT and Face StyleGAN models ($k=512$). The estimated PDF $\pi(\|\vec{v}\|^2_2)$ has heavier tails and differs significantly from the PDF of $\chi^2(k)$ for all three models, and thus invalidates the soap bubble effect argument exploited in PULSE.} %the Gaussian prior on the latent space $\mathcal{V}^+$ in PULSE.}
      \label{fig:gaussian_ansatz}
\end{figure}

After training the StyleGAN models, the validity of the Gaussian prior assumption  on the latent space $\mathcal{V}^+$---and thus of the norm constraint in the PULSE CSGM in Eq. \eqref{eq:pulse_original}---was investigated. The validation study was based on the following well-known theorem on standard Gaussian distributions \cite{vershynin2018random}:
\begin{theorem}\label{sec:theorem}
Let $\vec{a} \in \mathbb{R}^d$ be a standard Gaussian vector, i.e. $\vec{a} \sim \mathcal{N}(\vec{0},\vec{I}_d)$. Then $\|\vec{a}\|_2^2 \sim \chi^2(d)$, where $d$ is the degree of freedom of a $\chi^2-$distribution.
\end{theorem}
As a corollary to Theorem \ref{sec:theorem}, a necessary condition for a latent style vector $\vec{v}_i$ to follow a standard Gaussian distribution $\mathcal{N}(\vec{0},\vec{I}_{k})$ is that $\|\vec{v}_i\|_2^2 \sim \chi^2(k)$. With higher values of the degree of freedom $k$, the $\chi^2(k)$ distribution concentrates around the mode given by max$(0,k-2)$. 
To verify whether this condition holds true for the pre-trained MRI, CT and Face StyleGAN models, $10^7$ random realizations of a latent style vector $\vec{v} = \mathcal{T}G_m(\vec{z})$ were generated by sampling $\vec{z} \sim \mathcal{N}(\vec{0}, \vec{I}_k)$. 
The probability density function (PDF) of $\|\vec{v}\|^2_2$, denoted as $\pi(\|\vec{v}\|^2_2)$, was then estimated from those realizations. Figure \ref{fig:gaussian_ansatz} shows a comparison between $\pi(\|\vec{v}\|^2_2)$ and the PDF of the $\chi^2(k)$ distribution for the MRI, CT and Face StyleGAN models, for $k=512$. As expected, the $\chi^2(k)$ distribution is highly concentrated around the mode $k-2$, i.e. 510. 
On the contrary, the estimated PDF $\pi(\|\vec{v}\|^2_2)$ has much heavier tails and strongly differs from the PDF of $\chi^2(k)$ for all three StyleGAN models and hence it is evident that the soap bubble effect does not manifest in the $\mathcal{V}^+$ latent space.

In summary, this study invalidated the previously-held assumption that $\vec{v}_i \in \mathbb{R}^{k}$ is a standard Gaussian random vector, and presents a strong argument against constraining each $\vec{v}_i$ to lie on the spherical surface $S^{k-1}(\sqrt{k})$. Based on this finding, in the following section, an enhanced version of the PULSE method is proposed for empirical sampling. The proposed method, hereafter referred to as PULSE++, employs a more statistically principled approach to regularize the latent style vectors and thus encourages stronger data consistency in the alternate solutions.

\section{Generating multiple data-consistent solutions using PULSE++}
\label{sec:pulse++}
To account for the heavy tails of the estimated PDF $\pi(\|\vec{v}\|^2_2)$, PULSE++ replaces the projection step of $\vec{v}_i$ on the $S^{k-1}(\sqrt{k})$ sphere with a projection operation based onto the annular region $\mathcal{A} := \{\vec{v}_i \in \mathbb{R}^k | \delta_{min}\leq \|\vec{v}_i\|_2 \leq \delta_{max} \}$  while solving the CSGM optimization problem. The inner and outer radii $\delta_{min} < \sqrt{k} $ and $\delta_{max} > \sqrt{k}$ are chosen such that the probability of $\vec{v}_i$ lying inside $\mathcal{A}$ is equal to a pre-determined parameter $c \in (0,1)$. Specifically, $\delta_{min}$ and $\delta_{max}$ are chosen such that 
\begin{equation}\label{eq:hollow_ball}
    \Pi_{\rm emp}(\delta_{min}) = 1 - \Pi_{\rm emp}(\delta_{max}) = \frac{\gamma}{2},
\end{equation}
where $\Pi_{\rm emp}$ is the empirical cumulative distribution function (ECDF) of $\|\vec{v}_i\|_2$. The hyperparameter $\gamma$ can be chosen based on the desired trade-off between data consistency and the degree to which the alternate solutions are representative of the training distribution.
For a latent space vector $\vec{v} \in \mathbb{R}^k$, the projection operator $\mathcal{P}_{\mathcal{A}}: \mathbb{R}^k \mapsto \mathcal{A}$ is defined as 
\begin{equation}\label{eq:pulse++_projection}
    \mathcal{P}_{\mathcal{A}}(\vec{v}) = \begin{cases}
			\delta_{min}\dfrac{\vec{v}}{\|\vec{v}\|_2}, & \text{if $\|\vec{v}\|_2 < \delta_{min}$},\\
			\vec{v}, & \text{if $\delta_{min} \leq \|\vec{v}\|_2 \leq \delta_{max}$},\\
			\delta_{max}\dfrac{\vec{v}}{\|\vec{v}\|_2}, & \text{if $\|\vec{v}\|_2 > \delta_{max}$}.
		 \end{cases}
\end{equation}
The CSGM optimization problem in PULSE++ is stated as:
\begin{align}\label{eq:pulse_modified}
    \hat{\vec{V}},\hat{\boldsymbol{\Phi}} &= \argmin_{\vec{V},\boldsymbol{\Phi}} \mathcal{L}(\vec{V}, \boldsymbol{\Phi}):= \Big\{\mathcal{J}(\vec{g},\tilde{G}(\vec{V},\boldsymbol{\Phi})) + \mathcal{R}(\vec{V},\boldsymbol{\Phi})\Big\}, \nonumber\\
&s.t. \; \vec{v}_i \in \mathcal{A} \; \forall i \in \{1,\dots,L\},
\end{align}
where the regularization term is given by
\begin{equation}
    \mathcal{R}(\vec{V},\boldsymbol{\Phi}) = \lambda_c CROSS(\vec{V}) 
    +\frac{1}{2}\sum_{i=1}^L \|\boldsymbol{\phi}_i\|_2^2.
\end{equation}
Above, the pairwise Euclidean distance $CROSS(\vec{V}) \equiv \sum_{i=1}^{L-1} \sum_{j=i+1}^L \|\vec{v}_i-\vec{v}_j\|_2^2$\cite{menon2020pulse} is used in place of the $GEOCROSS$ term in the original PULSE formulation since latent vectors in $\mathcal{A}$ may have different norms. The hyperparameter $\lambda_c$ controls the strength of the $CROSS$-regularization to balance between ensuring consistency with the data and with the distribution of objects generated with the StyleGAN using latent vectors sampled from the original space $\mathcal{V}$. 

In a conventional CSGM formulation \cite{bora2017compressed}, regularization of the latent vectors for which the statistical distribution is known beforehand is performed by adding a penalty term that represents the negative log-probability of the density function. In PULSE, however, regularization on the Gaussian latent noise vectors in $\boldsymbol{\Phi}$ was performed by employing the soap bubble effect and imposing a strict norm constraint, resulting in an approximate Gaussian prior. In order to use the full knowledge of the prior distribution of $\boldsymbol{\Phi}$, the strict norm constraint on $\boldsymbol{\Phi}$ in Eq. \eqref{eq:pulse_original} was relaxed, and instead a penalty term was added in $ \mathcal{R}(\vec{V},\boldsymbol{\Phi})$ in the form of a negative log-probability density function of the Gaussian latent noise vectors in $\boldsymbol{\Phi}$.

Similar to Eq. \eqref{eq:pulse_original}, the optimization problem in PULSE++ is highly non-convex. Accelerated projected-gradient methods such as with the Adam optimizer can be employed to find alternate approximate solutions $\hcoeff \equiv \tilde{G}(\hat{\vec{V}},\hat{\boldsymbol{\Phi}})$ with multiple restarts of Eq. \eqref{eq:pulse_modified}. The estimates $\hat{\vec{V}}$ and $\hat{\boldsymbol{\Phi}}$ that minimize the objective function $\mathcal{L}(\vec{V}, \boldsymbol{\Phi})$ in Eq. \eqref{eq:pulse_modified} are obtained iteratively as follows. The iterates $\vec{V}^{[j]}$ and $\vec{\Phi}^{[j]}$ at iteration number $j$ are first updated using one step of the Adam algorithm to obtain $\vec{V}^{[j+\frac{1}{2}]}$ and $\vec{\Phi}^{[j+1]}$ respectively. Subsequently, the projection operator in Eq. \eqref{eq:pulse++_projection} is applied independently on each of the $L$ latent vectors that represent the columns of $\vec{V}^{[j+\frac{1}{2}]}$. The projection operations are performed in order to enforce the necessary constraint on the $\ell_2$-norm of each column of the latent matrix $\vec{V}^{[j+\frac{1}{2}]}$ that is defined by the annular region $\mathcal{A}$. For notational convenience, these $L$ independent projection operations will be simply denoted as $\mathcal{P}_{\mathcal{A}}(\vec{V}^{[j+\frac{1}{2}]})$. The Adam update followed by the projection operation $\mathcal{P}_{\mathcal{A}}(\vec{V}^{[j+\frac{1}{2}]})$ constitute a single projected-gradient step of the optimization procedure that produces the next iterates $\vec{V}^{[j+1]}$ and $\vec{\Phi}^{[j+1]}$. It should be noted that the objective function in Eq. \eqref{eq:pulse_modified} may not decrease monotonically with each such projected-gradient step\cite{reddi2019convergence}. Thus, the current best estimates $\hat{\vec{V}}$ and $\hat{\boldsymbol{\Phi}}$ of Eq. \eqref{eq:pulse_modified} are updated with the iterates $\vec{V}^{[j+1]}$ and $\vec{\Phi}^{[j+1]}$ only if $\mathcal{L}(\vec{V}^{[j+1]}, \boldsymbol{\Phi}^{[j+1]}) < \mathcal{L}(\hat{\vec{V}}, \hat{\boldsymbol{\Phi}})$. Finally, the approximate solution $\hcoeff = \tilde{G}(\hat{\vec{V}},\hat{\boldsymbol{\Phi}})$ is obtained when the maximum number of iterations is reached. The approximate solution $\hcoeff$ will be considered to be data-consistent if the data fidelity term $\mathcal{J}(\vec{g},\hcoeff)$ is less than a tolerance level $\epsilon_{\vec{n}}$ that is dependent on the measurement noise $\vec{n}$. The complete procedure for performing empirical sampling with PULSE++ is detailed in Algorithm \ref{pulse++_alg}.

\begin{algorithm}[!hbt]
\SetKw{KwInit}{Initialize}
\caption{Empirical sampling with PULSE++}\label{pulse++_alg}

\KwIn{Measurement data $\vec{g}$, forward operator $\mathcal{H}(\cdot)$, objective function $\mathcal{L}(\vec{V}, \boldsymbol{\Phi})$ from Eq. \eqref{eq:pulse_modified}; projection operator $\mathcal{P}_\mathcal{A}(\vec{V})$ from Eq. \eqref{eq:pulse++_projection}; annulus parameter $\gamma$, $CROSS$ parameter $\lambda_c$, learning rate $lr$ of Adam optimizer, number of gradient descent steps in Adam $n_{steps}$; number of alternate solutions $T$; acceptance tolerance $\epsilon_\vec{n}$}
\KwOut{Set $\bm{\hat{\vec{F}}}$ of data consistent object estimates}
$\bm{\hat{\vec{F}}} \leftarrow \{\}$\\
\For{$t\in\{1,\dots,T\}$}{
    \KwInit{$\vec{V}^{[1]}$ \emph{and} $\boldsymbol{\Phi}^{[1]}$ \emph{with} i.i.d. \emph{entries} $\sim \mathcal{N}(0,1)$}\\
    $\hat{\vec{V}} \leftarrow \vec{V}^{[1]}$; $\hat{\boldsymbol{\Phi}} \leftarrow \boldsymbol{\Phi}^{[1]}$\\
    \For{$j \in \{1, \ldots, n_{steps}\}$}{
        $\vec{V}^{[j+\frac{1}{2}]}, \boldsymbol{\Phi}^{[j+1]} \leftarrow \text{Adam}(\mathcal{L}(\vec{V}^{[j]},\boldsymbol{\Phi}^{[j]}))$\\
        $\vec{V}^{[j+1]} \leftarrow \mathcal{P}_{\mathcal{A}} (\vec{V}^{[j+\frac{1}{2}]})$\\
        \If{$\mathcal{L}(\vec{V}^{[j+1]}, \boldsymbol{\Phi}^{[j+1]}) < \mathcal{L}(\hat{\vec{V}}, \hat{\boldsymbol{\Phi}})$}{
            $\hat{\vec{V}} \leftarrow \vec{V}^{[j+1]}$; $\hat{\boldsymbol{\Phi}} \leftarrow \boldsymbol{\Phi}^{[j+1]}$
        }
    }
    $\hcoeff_t \leftarrow \tilde{G}(\hat{\vec{V}},\hat{\boldsymbol{\Phi}})$\\
    \If{$\mathcal{J}(\vec{g},\hcoeff_t) \leq \epsilon_{\vec{n}}$}{
        $\bm{\hat{\vec{F}}} \leftarrow \bm{\hat{\vec{F}}} \cup \{\hcoeff_t\}$
    }
}
\KwRet{$\bm{\hat{\vec{F}}}$}
\end{algorithm}

%\vspace{-0.4cm}
\section{Numerical studies}
\label{sec:Numerical studies}
Numerical studies were conducted to demonstrate the ability of the proposed PULSE++ method to produce multiple data-consistent solutions from the same tomographic measurements. Two stylized tomographic imaging systems were considered: one acquiring incomplete Fourier space measurements (Sec. \ref{sec:imaging_MRI}) and the other acquiring X-ray fan-beam CT projection data (Sec. \ref{sec:imaging_CT}). The advantage of PULSE++ over PULSE with respect to preserving data consistency is established from these numerical studies via an ablation study. The PULSE and PULSE++ methods were established by adapting an open-source implementation of PULSE for SISR in PyTorch \cite{pulse}. Since the training of a StyleGAN does not require any knowledge of the measurement process, the same pre-trained StyleGAN was employed for different sampling conditions in the PULSE and PULSE++ methods. The network architecture and pre-trained weights of MRI-StyleGAN and CT-StyleGAN were transferred from TensorFlow to PyTorch \cite{tf2torch}. 

For purposes of comparison, alternate solutions were also computed from incomplete Fourier space measurements by implementing a recently proposed approximate posterior sampling method \cite{jalal2021robust} that employs a score-based generative model (SGM) and annealed Langevin dynamics  \cite{song2019generative}. This approximate posterior sampling method will be referred to as SGM-based Langevin sampling (SGMLS) in the studies below. The SGMLS method requires training a state-of-the-art score-based generative model known as NCSNv2\cite{song2020improved}, which was performed by adapting an open-sourced implementation. The same training dataset of axial knee MRI images that was used to train MRI-StyleGAN was also employed to train the NCSNv2 model. The SGMLS method was performed by adapting a previous implementation \cite{langevin}. Similar to StyleGAN, the NCSNv2 model is also trained independent of the measurement process, and hence SGMLS was implemented with the same pre-trained NCSNv2 for different sampling conditions. The code repository for the numerical studies in our paper has been published \cite{pulse_pp}.

%\vspace{-0.2cm}
\subsection{Stylized imager that acquires incomplete Fourier space measurements}
\label{sec:imaging_MRI}
\begin{figure}
    \centering
    \includegraphics[width=\columnwidth]{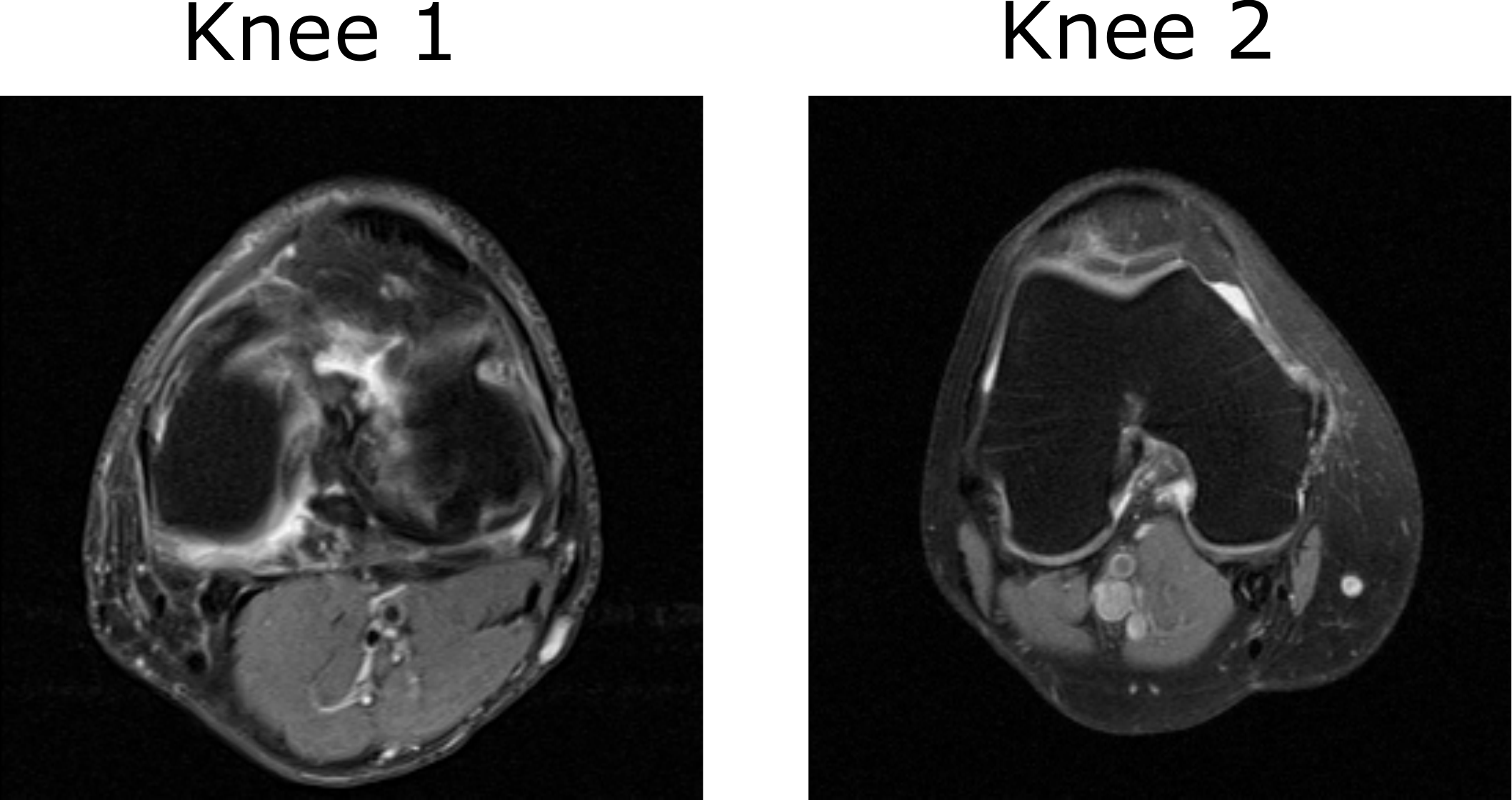}
    \caption{The objects Knee 1 and Knee 2 with size $256 \times 256$ from which noisy and incomplete k-space measurements were generated. Both the objects are displayed in the grayscale range of $[0,1]$.}
    \label{fig:mri_objects}
\end{figure}

A stylized imaging system that acquires incomplete 2D Fourier space (or k-space) measurements was considered. It should be noted that the goal of these preliminary studies was only to demonstrate and compare the performance of the proposed PULSE++ method against the original PULSE method and score-based posterior sampling, with respect to producing multiple data-consistent solutions from the same measurement data acquired under identical conditions. Hence, there is no attempt to model the real-world complexities of data-acquisition in MRI. Two different axial knee images of size $256 \times 256$ that belong to the NYU fastMRI dataset were considered to serve as objects $\vec{f}$, as shown in Fig. \ref{fig:mri_objects}. It should be noted that the images representing these objects, denoted as Knee 1 and Knee 2, were not included in the training dataset for the MRI-StyleGAN and the NCSNv2 models. Incomplete and noisy k-space data were simulated from these two objects for demonstrating the numerical studies. The acceleration factor was defined as $R=N/M > 1$, where $M$ denotes the number of k-space samples measured and $N$ is the dimension of the object.
The imaging operator was modeled as $\mathcal{H} = \vec{M}\mathcal{F}$, where $\mathcal{F}$ denotes the 2D Fast Fourier Transform (FFT) and the binary matrix $\vec{M} \in \{0,1\}^{M \times N}$ represents a random Cartesian k-space sampling mask. The measurement noise $\vec{n} \in \mathbb{E}^M$ was sampled from an i.i.d complex Gaussian distribution $\mathcal{CN}(\vec{0},\sigma^2\vec{I}_M)$\cite{aja2016statistical}. In the numerical experiments, the size of the image was set to $N=256^2$. Two different acceleration factors $R=6,8$ and two levels of noise $\sigma=0.07, 0.05$ were investigated. The random Cartesian sampling masks were generated using open-sourced codes \cite{fastmri}. The MRI-StyleGAN model described in Sec. \ref{sec:validation} was employed for performing empirical sampling with PULSE and PULSE++ using the same k-space data for each combination of $\sigma$ and $R$. Since the measurement noise was i.i.d. Gaussian, the data fidelity term  was specified as $\mathcal{J}(\vec{g},\hcoeff) = \frac{1}{2\sigma^2} \|\vec{g}-\mathcal{H}\hcoeff\|^2_2$, where $\vec{g} \in \mathbb{R}^M$ is the k-space data and $\hcoeff \in \mathbb{R}^N$ denotes the estimated object. The tolerance level $\epsilon_{\vec{n}}=\frac{M}{2}$ for accepting the data-consistent alternate solution $\hcoeff$ was determined based on Morozov's discrepancy principle \cite{morozov1966solution}. The learning rate $lr$ of the Adam optimizer in the CSGM formulation for the PULSE and PULSE++ methods was set as 0.4, similar to the original implementation of PULSE in \cite{menon2020pulse}. The number of gradient descent steps $n_{steps}$ performed to obtain each alternate solution was 10,000. The initial step size $\eta$ for annealed SGMLS \cite{song2019generative} in the score-based posterior sampling method was selected within a range $[10^{-8},10^{-6}]$ for each numerical experiment to prevent the divergence of the data fidelity term or generation of extremely noisy alternate solutions\cite{ma2022accelerating}. An alternate solution obtained using either PULSE or PULSE++ was completed in $\sim$9 minutes, while each alternate solution from the score-based posterior sampling method using the NCSNv2 model was computed in $\sim$7 minutes, with all the experiments performed using an NVIDIA 1080 Ti GPU.

\subsection{CT imaging system with limited angular range}
\label{sec:imaging_CT}
\begin{figure}
    \centering
    \includegraphics[width=\columnwidth]{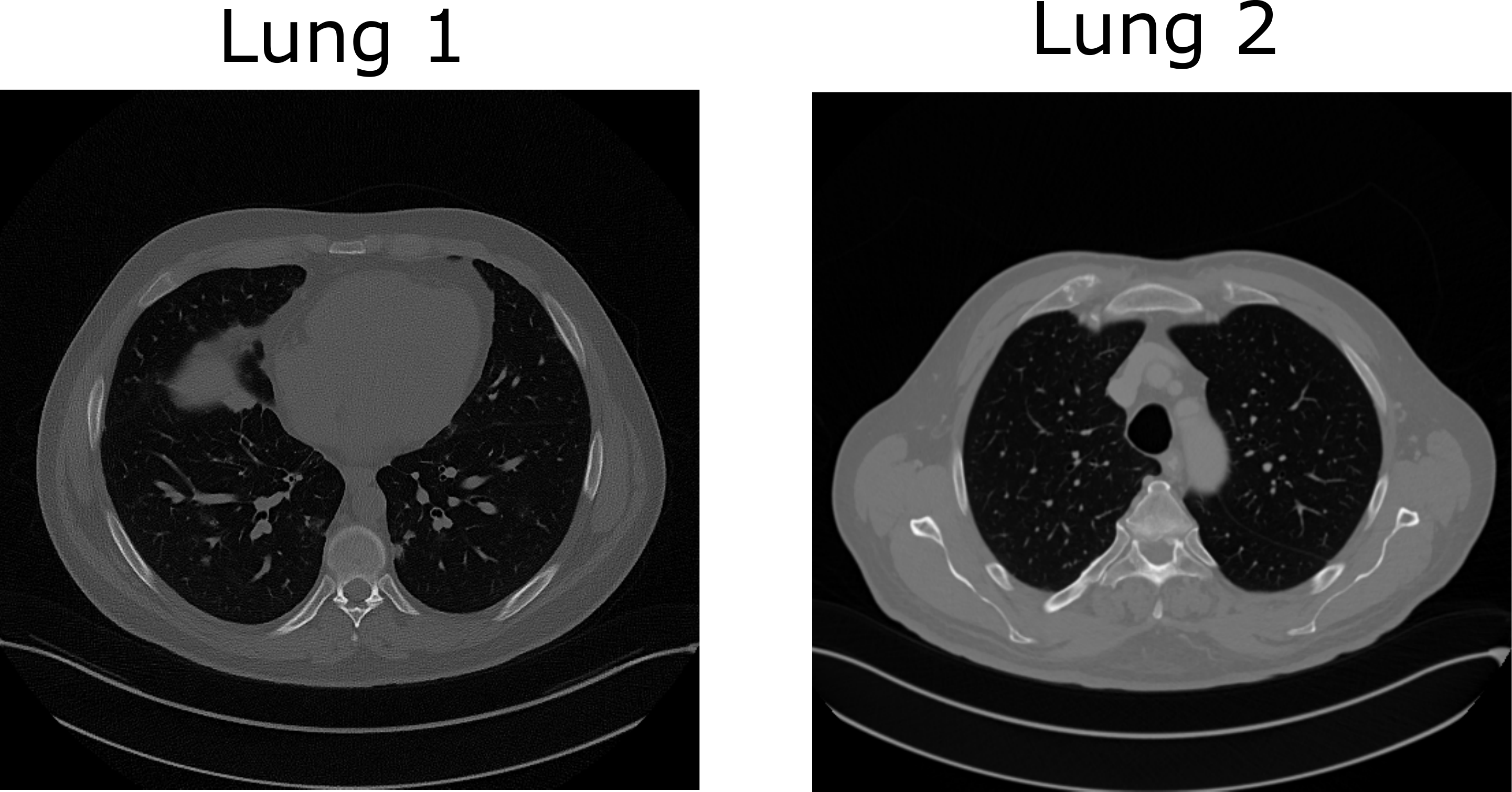}
    \caption{The objects Lung 1 and Lung 2 with size $512 \times 512$ from which noisy and incomplete X-ray projection data were generated. Both the objects are displayed in the grayscale range of $[0,1]$.}
    \label{fig:ct_objects}
\end{figure}

Numerical studies were also performed with incomplete and noisy X-ray CT measurements from a stylized CT imaging system. The objective of these studies was to demonstrate the ability of the PULSE++ method to perform empirical sampling in higher-dimensional spaces (resolution $512 \times 512$ pixels) in a computationally feasible manner. The CT-StyleGAN model introduced in Sec. \ref{sec:validation} was employed for these studies to find alternate solutions from the same X-ray photon projection data. Two separate CT lung images of size $512 \times 512$ pixels were selected from the NIH DeepLesion dataset to represent objects $\vec{f}$ from which measurement data were simulated. It should be noted that these images were not included during training of the StyleGAN. The objects, denoted as Lung 1 and Lung 2, are shown in Fig. \ref{fig:ct_objects}. The maximum linear attenuation coefficient values in Lung 1 and Lung 2 were 0.063 mm$^{-1}$ and 0.046 mm$^{-1}$, respectively. The physical unit of each pixel (px) was 0.82 mm \cite{yan2018deeplesion}. A fan-beam geometry with a linear detector array and a monoenergetic source was assumed. Projection data were simulated for 120 views spanning the limited angular range $[0^{\circ},119^{\circ}]$. The noiseless X-ray measurements $\bar{\vec{g}} \in \mathbb{R}^M$ from an object $\coeff \in \mathbb{R}^N$ were modeled as \cite{ding2016modeling}
\begin{equation}\label{eq:photons}
    \bar{\vec{g}} = I_0 \exp(-\vec{Hf}),
\end{equation}
where the system matrix $\vec{H} \in \mathbb{E}^{M \times N}$ is the fan-beam projector and $I_0$ is the intensity of an unattenuated beam. The fan-beam projector $\vec{H}$ was implemented using the Air Tools II library \cite{hansen2018air}. The noisy intensity measurements $\vec{g} \in \mathbb{R}^M$ were Poisson-distributed with a mean of $\bar{\vec{g}}$\cite{leng2019photon}. Higher values of $I_0$ result in a higher signal-to-noise ratio (SNR) in the intensity measurements. Since the aim of this simulation study was primarily to assess the ability of the PULSE++ method to perform empirical sampling with high-dimensional objects, additional physical factors required to accurately model a real-world CT imaging system such as beam spectrum, photon scattering and dark current effects were not considered. Numerical studies were conducted using simulated projection data with values of $I_0=10^3$ and $I_0=10^5$, which correspond to measurement data that have different levels of photon noise. The data fidelity term $\mathcal{J}(\vec{g},\hcoeff)$ in Eq. \eqref{eq:pulse_modified} was defined as the Kullback-Leibler (KL) divergence between the noisy measurement data $\vec{g}$ and the forward projection data $\hat{\vec{g}}=I_0 \exp(-\vec{H}\hcoeff)$ \cite{kak2002principles}. Unlike Gaussian noise, however, the exact tolerance level $\epsilon_\vec{n}$ required to apply Morozov's discrepancy principle is not explicitly available as Poisson-distributed noise cannot be marginalized\cite{sixou2018morozov}. A heuristic rule was employed for accepting an alternate solution, stipulated as 
\begin{equation}\label{eq:poisson_dp}
    \mathcal{J}(\vec{g},\hcoeff) \leq \mathcal{J}(\vec{g},\tilde{\coeff}) := \epsilon_{\vec{n}},
\end{equation}
where $\tilde{\coeff} = \tilde{G}(\tilde{\vec{V}}, \tilde{\boldsymbol{\Phi}})$ is the Euclidean projection (embedding) of the measured object $\coeff$ onto the range of the generator network $\tilde{G}$. Such a heuristic rule was designed to eliminate the contribution to data inconsistency due to representation error\cite{bora2017compressed} in the StyleGAN. Here, representation error refers to the minimum Euclidean distance of an object from the true data distribution with respect to the object distribution embodied by the generator network of the StyleGAN. 
Each CSGM run of the PULSE++ method with the CT measurement data completed in $\sim 14$ minutes on an NVIDIA V100 GPU.

\subsection{Ablation study}
\label{sec:ablation}
An ablation study was performed to critically assess the impact of each of the enhancements introduced on the latent variables $\vec{V}$ and $\boldsymbol{\Phi}$ to develop the PULSE++ method from the baseline PULSE method. Intermediate methods that represent each of these enhancements were implemented as described below:
\subsubsection{(PULSE$_1$) PULSE + projection operator $\proj$}
\label{sec:pulse_1}
The inaccurate strict norm constraint $\vec{v}_i \in S^{k-1}(\sqrt{k})$ in Eq. \eqref{eq:pulse_original} is replaced with the constraint $\vec{v}_i \in \mathcal{A}$, where $\mathcal{A}$ is the annulus region defined using the estimated PDF $\pi(\|\vec{v}\|^2_2)$ in Sec. \ref{sec:pulse++}:
\begin{align}\label{eq:pulse_1}
    \hat{\vec{V}},\hat{\boldsymbol{\Phi}} &= \argmin_{\vec{V},\boldsymbol{\Phi}} \mathcal{L}(\vec{V}, \boldsymbol{\Phi}) 
    := \Big\{\mathcal{J}(\vec{g},\tilde{G}(\vec{V},\boldsymbol{\Phi})) + \mathcal{R}(\vec{V})\Big\}, \nonumber\\
    &s.t. \; \vec{v}_i \in \mathcal{A}, \;
    \boldsymbol{\phi}_i \in S^{p_i-1}(\sqrt{p_i}) \: \forall i \in \{1,\dots,L\},
    \end{align}
    where the regularization term is given by
    \begin{equation}
        \mathcal{R}(\vec{V}) = \lambda_c CROSS(\vec{V}). 
    \end{equation}

\subsubsection{(PULSE$_2$) PULSE + log-likelihood penalty on $\Phi$}
\label{sec:pulse_2}
The constraint $\boldsymbol{\phi}_i \in S^{p_i-1}(\sqrt{p_i})$ denoting an approximate Gaussian prior in PULSE is substituted with the statistically consistent log-probability density function penalty:
\begin{align}\label{eq:pulse_2}
    \hat{\vec{V}},\hat{\boldsymbol{\Phi}} &= \argmin_{\vec{V},\boldsymbol{\Phi}} \mathcal{L}(\vec{V}, \boldsymbol{\Phi})
    := \Big\{\mathcal{J}(\vec{g},\tilde{G}(\vec{V},\boldsymbol{\Phi})) + \mathcal{R}(\vec{V},\boldsymbol{\Phi})\Big\}, \nonumber\\
    &s.t. \; \vec{v}_i \in S^{k-1}(\sqrt{k}) \: \forall i \in \{1,\dots,L\},
\end{align}
where the regularization term is given by:
\begin{equation}
    \mathcal{R}(\vec{V},\boldsymbol{\Phi}) = \lambda_g GEOCROSS(\vec{V}) 
    +\frac{1}{2}\sum_{i=1}^L \|\boldsymbol{\phi}_i\|_2^2. 
\end{equation}
Finally, combining both the enhancements in the two intermediate methods PULSE$_1$ and PULSE$_2$ results in the PULSE++ method described by Eq. \eqref{eq:pulse_modified}. 

%\newpage
\section{Results}
\label{sec:results}
\subsection{Empirical sampling from Fourier space measurements}
\label{sec:results_mri}
\subsubsection{Visual assessment}
\label{sec:mri_visual}
\begin{figure}[!htbp]
    \centering
     \includegraphics[width=0.96\linewidth]{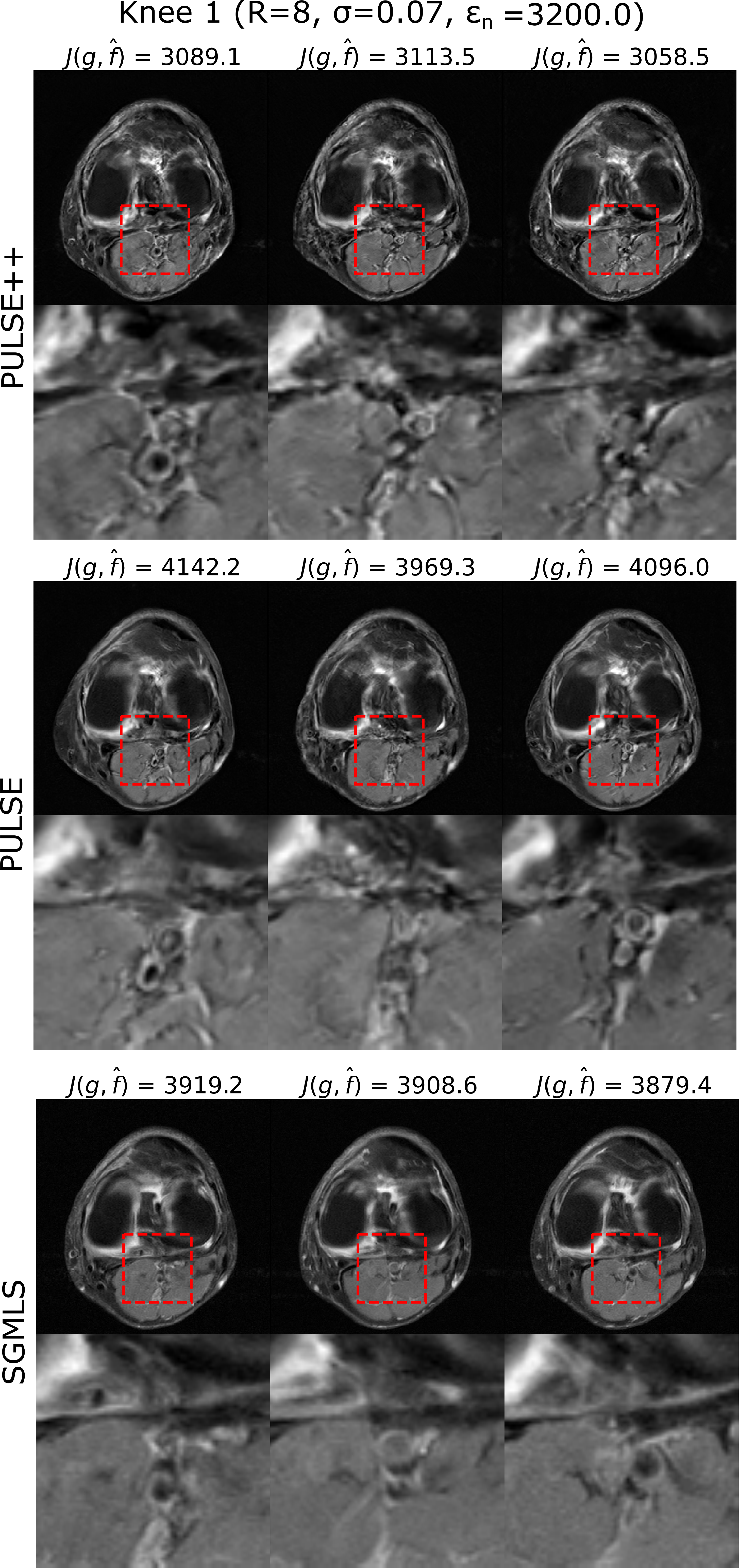}
      \caption{Samples of alternate solutions $\hcoeff$ obtained using PULSE++, PULSE and score-based posterior sampling from the same k-space data $\vec{g}$ produced from Knee 1 for $R=8$ and $\sigma=0.07$, with the data consistency tolerance $\epsilon_{\vec{n}}=3200.0$. The data fidelity value $\mathcal{J}(\vec{g},\hcoeff)$ for each alternate solution $\hcoeff$ is provided in the respective figure headings. Zoomed-in images of the same region indicated by a red bounding box are shown below each alternate solution that demonstrate distinct structures. However, only the alternate solutions produced by PULSE++ have data fidelity values lower than $\epsilon_{\vec{n}}$, while the alternate solutions obtained using PULSE and score-based posterior sampling are not data-consistent. All the alternate solutions are displayed in the grayscale range $[0,1]$.}
      \label{fig:mri_8x}
\end{figure}

Samples of alternate solutions generated by PULSE++ ($\gamma=0.001, \lambda_c=0.01$), PULSE ($\lambda_g=0.1$) and SGMLS ($\eta=2\times10^{-7}$) from the same k-space data produced by Knee 1, corresponding to $R=8$ and $\sigma=0.07$, are shown in Fig. \ref{fig:mri_8x}. For each method, the alternate solutions exhibit considerable diversity while being produced by use of the same measurement data. Additional examples of alternate solutions obtained with the PULSE++, PULSE and SGMLS methods are provided in the form of supplementary material as described in the Appendix. However, among the three methods, only the alternate solutions produced by the PULSE++ method satisfied the stipulated data-consistency criterion based on Morozov's discrepancy principle. This establishes an advantage of the PULSE++ method over PULSE in achieving the desired data consistency by properly accounting for the heavy tails of the empirical distribution of the latent style vectors in $\mathcal{V}^+$. This also demonstrates the superiority of PULSE++ over SGMLS in preserving data consistency of alternate solutions.

\begin{figure}[!htbp]
    \centering
     \includegraphics[width=\linewidth]{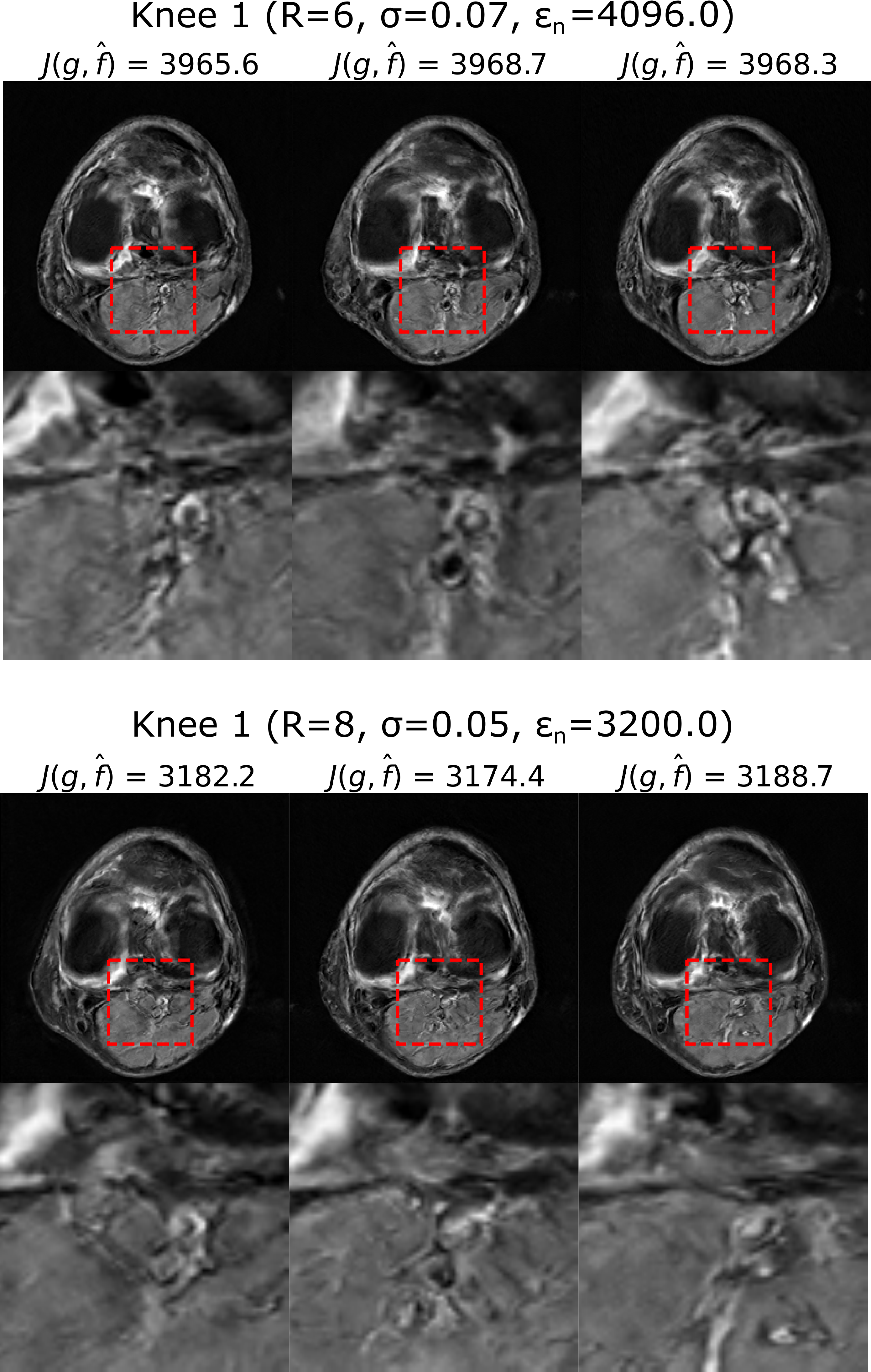}
      \caption{Alternate data-consistent solutions obtained using PULSE++ from k-space data produced by Knee 1 for different sampling conditions \{$R=6$, $\sigma=0.07$\} (top) and \{$R=8$, $\sigma=0.05$\} (bottom) using the same MRI-StyleGAN model as in Fig. \ref{fig:mri_8x}. Zoomed-in images of the same region in the alternate solutions indicated by a red bounding box show distinct structures. The alternate solutions are displayed in the grayscale range $[0,1]$.}
      \label{fig:mri_diff_params}
\end{figure}

\begin{figure}[!htbp]
    \centering
     \includegraphics[width=\linewidth]{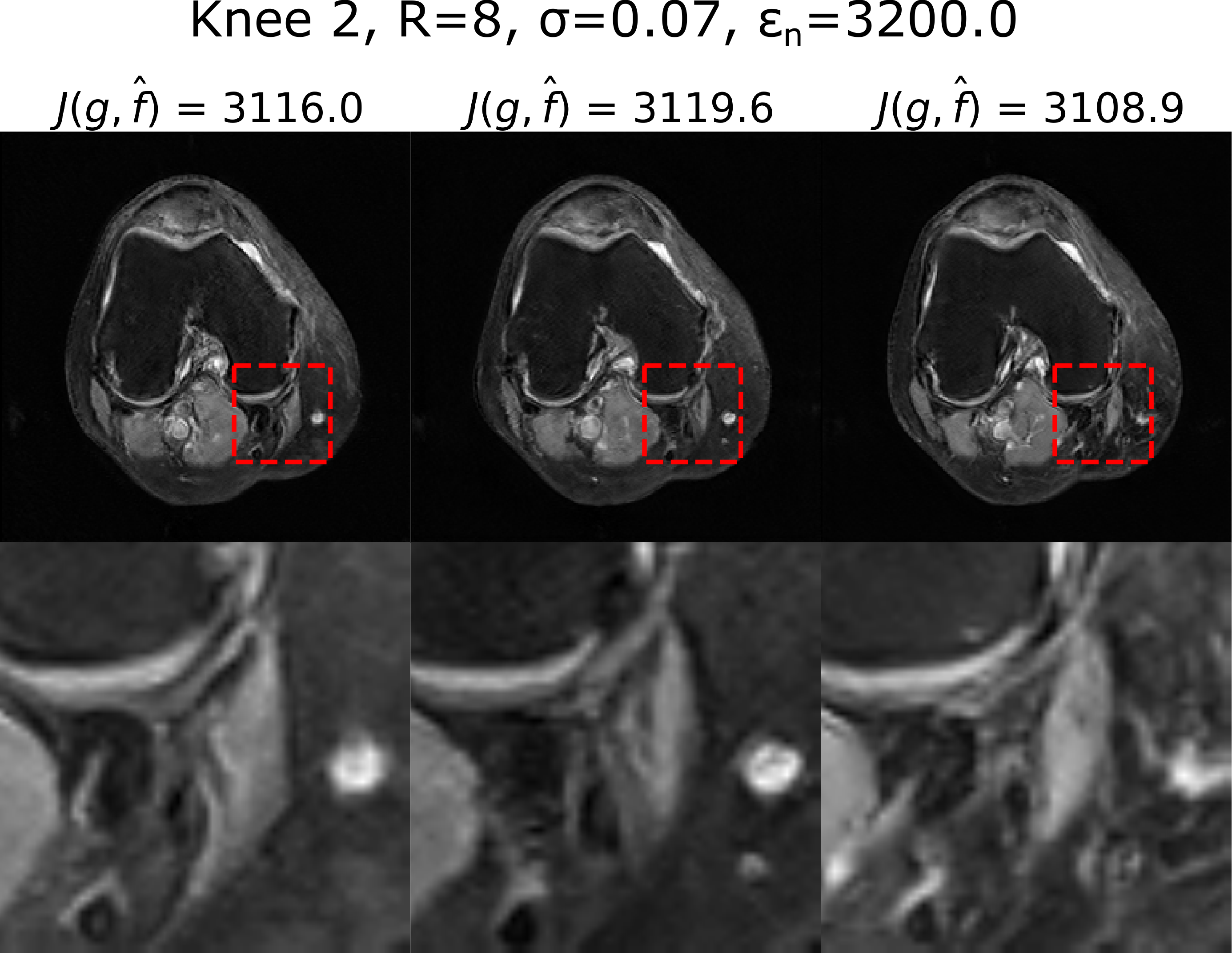}
      \caption{PULSE++ can produce alternate data-consistent solutions for k-space data from different objects within the same distribution on which the StyleGAN is trained, as shown here for Knee 2 using the MRI-StyleGAN model with $R=8$ and $\sigma=0.07$. Both the objects Knee 1 and Knee 2 belong to the NYU fastMRI dataset, but were excluded from the training set of the MRI-StyleGAN. Zoomed-in images of the same region in the alternate solutions indicated by a red bounding box show distinct structures. The alternate solutions are displayed in the grayscale range $[0,1]$.}
      \label{fig:mri_knee2}
\end{figure}

Figure \ref{fig:mri_diff_params} shows samples of data-consistent alternate solutions obtained with PULSE++ from k-space produced by Knee 1 when the sampling pattern or the noise level is varied, e.g. with sampling conditions \{$R=6$, $\sigma=0.07$\} and \{$R=8$, $\sigma=0.05$\}. The corresponding PULSE++ parameters were \{$\gamma=10^{-4}$, $\lambda_c=0.01$\} and \{$\gamma=10^{-5}$, $\lambda_c=0.001$\} respectively. Additionally, using the same pre-trained StyleGAN model, data-consistent alternate solutions were obtained with PULSE++ \{$\gamma=0.01$, $\lambda_c=0.001$\} from k-space data with $R=8$ and $\sigma=0.07$ corresponding to Knee 2 (Fig. \ref{fig:mri_objects}), as shown in Fig. \ref{fig:mri_knee2}.

\subsubsection{Ablation study}
\label{sec:ablation_results}
An ablation study was performed to comprehensively assess the improvement in data consistency yielded by the PULSE++ method, as outlined in Sec. \ref{sec:ablation}, using alternate solutions from the k-space data corresponding to $R=8$ and $\sigma=0.07$. For evaluation, 100 alternate solutions were computed using each of the methods PULSE ($\lambda_g=0.1$), $\text{PULSE}_1 (\gamma=0.001, \lambda_c=0.01)$, $\text{PULSE}_2 (\lambda_g=0.1)$ and PULSE++ ($\gamma=0.001, \lambda_c=0.01$). A box plot of the data fidelity values obtained with each method is shown in Fig. \ref{fig:box_plot}. It was observed that as high as 92\% of the alternate solutions obtained with the $\text{PULSE}_1$ method satisfied the data fidelity tolerance $\epsilon_{\vec{n}}$, whereas none of the alternate solutions produced by the original PULSE approach had a data fidelity value below $\epsilon_{\vec{n}}$. This clearly demonstrates the impact of utilizing improved statistical assumptions about the $\mathcal{V}^+$ space by use of the projection operator $\mathcal{P}_{\mathcal{A}}(\vec{v})$. On the other hand, introducing only the log-probability density penalty term for the Gaussian noise latent vectors in $\boldsymbol{\Phi}$, as in $\text{PULSE}_2$, resulted in similar data fidelity values on average as compared to PULSE, but with a lower variance. 
%Evidently, incorporating the soap bubble effect on $\boldsymbol{\Phi}$ in PULSE and thus using an approximate prior, as compared to the conventional log-probability density penalty, does not impact data consistency to the same extent as the introduction of $\mathcal{P}_{\mathcal{A}}(\vec{v})$ in the style latent space $\mathcal{V}^+$. 
Empirically, there is no significant benefit to preserving data consistency by imposing an approximate Gaussian prior with a strict norm constraint on $\Phi$ as performed in the PULSE method. Finally, it was observed that 93\% of alternate solutions produced by the PULSE++ method satisfied the data fidelity tolerance. Thus, the ablation study demonstrates the impact on achieving data consistency due to the individual enhancements introduced in the PULSE method to formulate the PULSE++ approach.

\subsubsection{Comparison of data consistency against SGMLS}
\begin{figure}[!htbp]
    \centering
     \includegraphics[width=\linewidth]{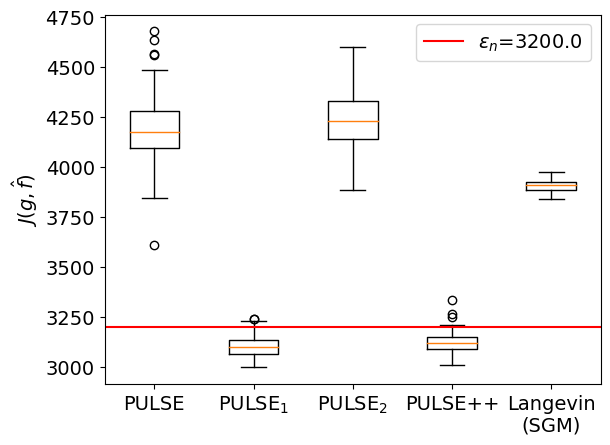}
      \caption{A box plot of data fidelity values of alternate solutions obtained with different methods from the k-space corresponding to Knee 1 with $R=8$ and $\sigma=0.07$. For each method, data fidelity values of 100 alternate solutions were plotted. The methods include the original PULSE method, PULSE with only the projection operator $\mathcal{P}_{\mathcal{A}}(\vec{v})$ for the style latent space $\mathcal{V}^+$ ($\text{PULSE}_1$), PULSE with only the log-likelihood penalty term on the noise latent vectors in $\boldsymbol{\Phi}$ in place of the spherical projection to impose the Gaussian prior ($\text{PULSE}_2$), PULSE++, and SGMLS. The plot demonstrates the impact on data fidelity obtained with each of the individual enhancements introduced in PULSE to produce the PULSE++ method. It is evident that utilization of improved statistical assumptions about the style latent space $\mathcal{V}^+$ contributes significantly to ensure that the data fidelity tolerance $\epsilon_{\vec{n}}$ is satisfied. Additionally, the plot establishes the superiority of PULSE++ over SGMLS in terms of preserving data consistency.} 
      \label{fig:box_plot}
\end{figure}

The data fidelity from 100 samples obtained using SGMLS is shown in the box plot in Fig. \ref{fig:box_plot}. The data fidelity values of the SGMLS samples significantly exceed the tolerance level $\epsilon_{\vec{n}}$. This observation demonstrates that SGMLS may be insufficient for achieving alternate solutions that also preserve data consistency.

\subsubsection{Uncertainty quantification}
\label{sec:mri_uq}
Uncertainty quantification was performed from the alternate solutions obtained with the PULSE, PULSE++ and SGMLS methods for each set of k-space data corresponding to different sampling conditions. The \textit{uncertainty map} $\cUM$ was computed as the pixel-wise standard deviation of the alternate solutions $\{\hcoeff_t\}_{t=1}^T$, where $T$ is the number of alternate solutions. Additionally, uncertainty maps were computed separately for the measurable and null space components \cite{barrett2013foundations} of the alternate solutions. The measurable and null space component of an object $\coeff \in \mathbb{R}^N$ in the domain of system matrix $\vec{H}\in \mathbb{E}^{M \times N}$ are defined as $\cmeas = \vec{H}^+\vec{H}\coeff$ and $\cnull = [\vec{I}_N-\vec{H}^+\vec{H}]\coeff$ respectively, where $\vec{H}^+$ is the Moore-Penrose pseudoinverse of $\vec{H}$. The null space component $\cnull$ is ``invisible'' to $\vec{H}$, and only $\cmeas$ contributes to the forward projection data $\vec{H}\coeff$. The uncertainty maps of the measurable and null space components of alternate solutions were denoted as $\cUMmeas$ and $\cUMnull$ respectively. If multiple solutions are consistent with the same measurement data, it is expected that the variability expressed by $\cUMnull$ will be higher than that expressed by $\cUMmeas$.

\begin{figure}[!htb]
    \centering
     \includegraphics[width=\linewidth]{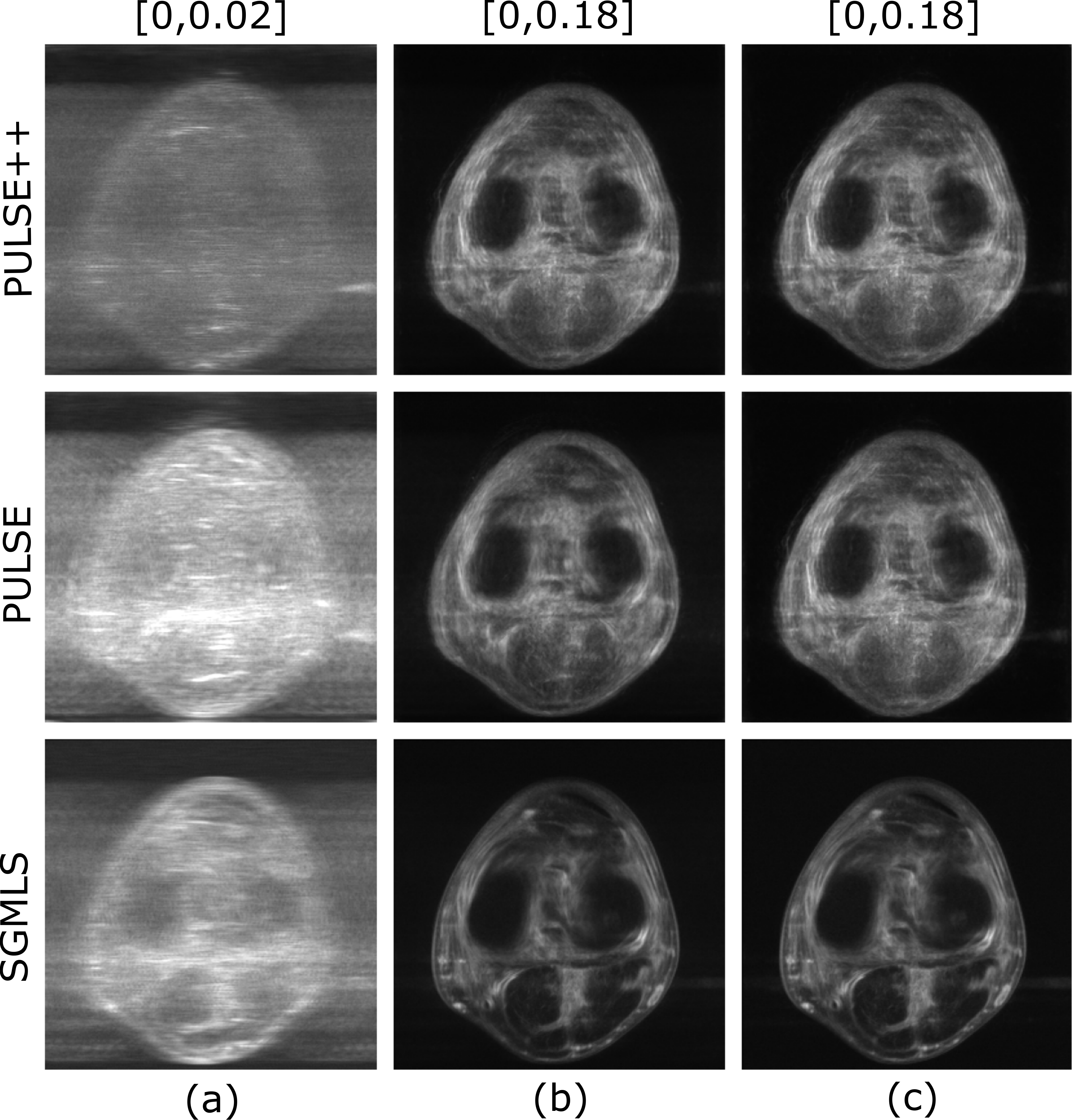}
      \caption{Uncertainty maps (a) $\cUMmeas$, (b) $\cUMnull$ and (c) $\cUM$ using PULSE++, PULSE and SGMLS methods from the same k-space data produced by Knee 1 with $R=6$ and $\sigma=0.07$. The grayscale range of each type of uncertainty map is indicated at the top of the corresponding column. It can be observed that the uncertainty primarily lies in the null space component for all methods. The PULSE++ method has significantly lower uncertainty in the measurable component compared to both PULSE and SGMLS methods, indicating enhanced data consistency in alternate solutions produced by PULSE++. The uncertainty estimated in the null space component and total uncertainty in PULSE++ is noticeably higher compared to SGMLS. This indicates that SGMLS fails to explore the manifold of data-consistent alternate solutions and underestimates uncertainty.}
      \label{fig:mri_uncertainty_maps}
\end{figure}

From each set of k-space data, uncertainty maps were computed from $T=100$ alternate solutions obtained with each of the three methods. The uncertainty maps corresponding to Knee 1 and system parameters $R=6$ and $\sigma=0.07$ are shown in Fig. \ref{fig:mri_uncertainty_maps}. For all three methods, it was observed that the diversity in the alternate solutions was primarily due to variations in the null space component. As anticipated, the uncertainty in the measurable component of the alternate solutions obtained with the PULSE method was considerably higher as compared to PULSE++ due to a less accurate projection step. Furthermore, the SGMLS method also yielded higher uncertainty in the measurable component and significantly less uncertainty in the null space component as compared to PULSE++. Combined with the reduced bias in data fidelity as observed in Fig. \ref{fig:box_plot}, the uncertainty maps of the measurable component further illustrate the ability of PULSE++ to enforce stronger data consistency among alternate solutions as compared to PULSE and SGMLS, while still maintaining high variability in the null space component.

Additionally, three figures-of-merit (FOMs) were computed from the uncertainty maps to characterize the degree of variability associated with the alternate solutions obtained via each method. The \textit{total uncertainty} FOM $\| \cUM \|^2_2$ is the total estimated variance from all the pixels in an alternate solution. Similarly, the uncertainty FOMs associated with the measurable and null space components of alternate solutions are $\| \cUMmeas \|^2_2$ and $\| \cUMnull \|^2_2$ respectively. It should be noted that, in theory, $\| \cUM \|^2_2 = \| \cUMmeas \|^2_2 + \| \cUMnull \|^2_2$ since $\hcmeas$ and $\hcnull$ are orthogonal to each other. However, there may be small discrepancies between those quantities due to floating point arithmetic and numerical approximations in the iterative computation of $\hcmeas$ and $\hcnull$\cite{barrett2013foundations}. Table \ref{tab:mri_reconstruction_risk} summarizes the uncertainty FOMs of the alternate solutions and their measurable and null space components, corresponding to PULSE++, PULSE and SGMLS methods for objects Knee 1 and Knee 2 with different system parameter settings. In all cases, the uncertainty FOM in the null space component was significantly higher than compared to that in the measurable component. This is consistent with the uncertainty maps in Fig. \ref{fig:mri_uncertainty_maps}. As expected, when the acceleration factor $R$ was increased, there was a decrease in the  uncertainty FOM in the measurable component while the uncertainty FOM in the null space component increased. Predictably, a reduction in the noise level $\sigma$ for the same value of $R$ also diminished the uncertainty FOM in the measurable component of the alternate solutions. Furthermore, for the same acceleration factor $R$ and noise level $\sigma$, both PULSE and SGMLS methods possessed a consistently higher uncertainty FOM in the measurable component as compared to PULSE++. This corroborates that the PULSE++ method significantly reduces the risk of data inconsistency for generating alternate solutions. On the other hand, the total uncertainty estimated by the SGMLS method is noticeably lower compared to both PULSE++ and PULSE. This suggests that SGMLS fails to explore the manifold of data-consistent alternate solutions and thus underestimates uncertainty.
\begin{table}[!htbp]
    \centering
    \caption{Summary of uncertainty FOMs of alternate solutions from the same k-space data for different values of $R$ and $\sigma$}
    \resizebox{\columnwidth}{!}{
    \begin{tabular}{c|lcc|ccc}
    \toprule
     Object & Method & $R$ &$\sigma$ &$\| \cUMmeas \|^2_2$ & $\| \cUMnull \|^2_2$ &$\| \cUM \|^2_2$\\
     \midrule
     \multirow{9}{*}{Knee 1}
     &PULSE++ &6  &0.07  & 3.74  & 152.19    & 155.93 \\
     &PULSE   &6  &0.07  & 9.61  & 133.05    & 142.66 \\
     &SGMLS   &6  &0.07  & 6.24  & 70.33    & 76.57 \\
     \cmidrule{2-7}
     &PULSE++ &8  &0.07  & 3.46  & 196.48    & 199.95 \\
     &PULSE   &8  &0.07  & 6.55  & 163.13    & 169.68 \\
     &SGMLS   &8  &0.07  & 5.34  & 84.07    & 89.41 \\
     \cmidrule{2-7}
     &PULSE++ &8  &0.05  & 2.26  & 184.11    & 186.37 \\
     &PULSE   &8  &0.05  & 5.77  & 149.01    & 154.77 \\
     &SGMLS   &8  &0.05  & 4.30  & 90.16    & 94.46 \\
     \midrule
     \midrule
     
     \multirow{9}{*}{Knee 2}
     &PULSE++ &6  &0.07  & 2.67  & 76.22    & 78.89 \\
     &PULSE   &6  &0.07  & 3.87  & 60.36    & 64.23 \\
     &SGMLS   &6  &0.07  & 4.86  & 48.11    & 52.96 \\
     \cmidrule{2-7}
     &PULSE++ &8  &0.07  & 2.44  & 101.77    & 104.20 \\
     &PULSE   &8  &0.07  & 3.25  & 83.09    & 86.35 \\
     &SGMLS   &8  &0.07  & 4.24  & 55.93    & 60.17 \\
     \cmidrule{2-7}
     &PULSE++ &8  &0.05  & 1.58  & 92.84    & 94.42 \\
     &PULSE   &8  &0.05  & 2.62  & 69.76    & 72.38 \\
     &SGMLS   &8  &0.05  & 3.51  & 61.87  & 65.38 \\
     \bottomrule
    \end{tabular}}
    \label{tab:mri_reconstruction_risk}
\end{table}

%\vspace{-0.5cm}
\subsection{Empirical sampling from limited-angle CT measurements}
\label{sec:results_ct}

\subsubsection{Visual assessment}
\label{sec:ct_visual}
\begin{figure}[!htb]
    \centering
     \includegraphics[width=0.95\linewidth]{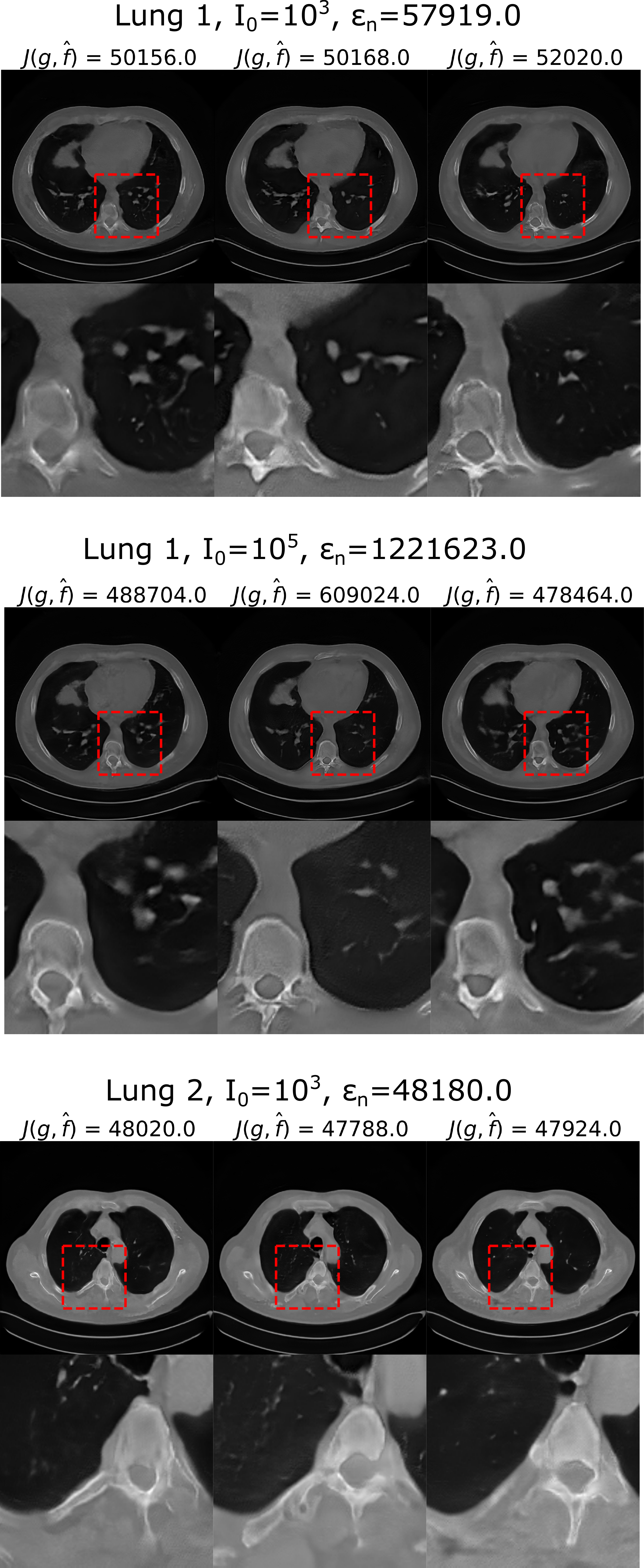}
     %  \vspace{-0.8cm}
      \caption{Alternate data-consistent solutions obtained using the PULSE++ method using the CT-StyleGAN model under different settings, for projection data from Lung 1 ($I_0=10^3$, $I_0=10^5$) and Lung 2 ($I_0=10^3$). Zoomed-in images of the same region in the alternate solutions inside the red bounding box demonstrate diversity in a number of fine-scale structures. The grayscale range of the alternate solutions is [0,1].}
      \label{fig:ct_pulse++}
\end{figure}

Samples of data-consistent alternate solutions obtained with the PULSE++ method using the same CT-StyleGAN model are shown in Fig. \ref{fig:ct_pulse++}, corresponding to limited-angle projection data from Lung 1 ($I_0=10^3$, $I_0=10^5$) and Lung 2 ($I_0=10^3$). Additional alternate solutions are provided in the supplementary material described in the Appendix. The alternate solutions in each case displayed considerable variability in fine-scale structures. This illustrates the ability of the proposed PULSE++ method to produce diverse data-consistent solutions from the same measurement data for high-dimensional objects, which may be computationally infeasible with currently available posterior sampling methods.

\subsubsection{Uncertainty quantification}
\label{sec:ct_uq}
\begin{figure}[!htb]
    \centering
     \includegraphics[width=\linewidth]{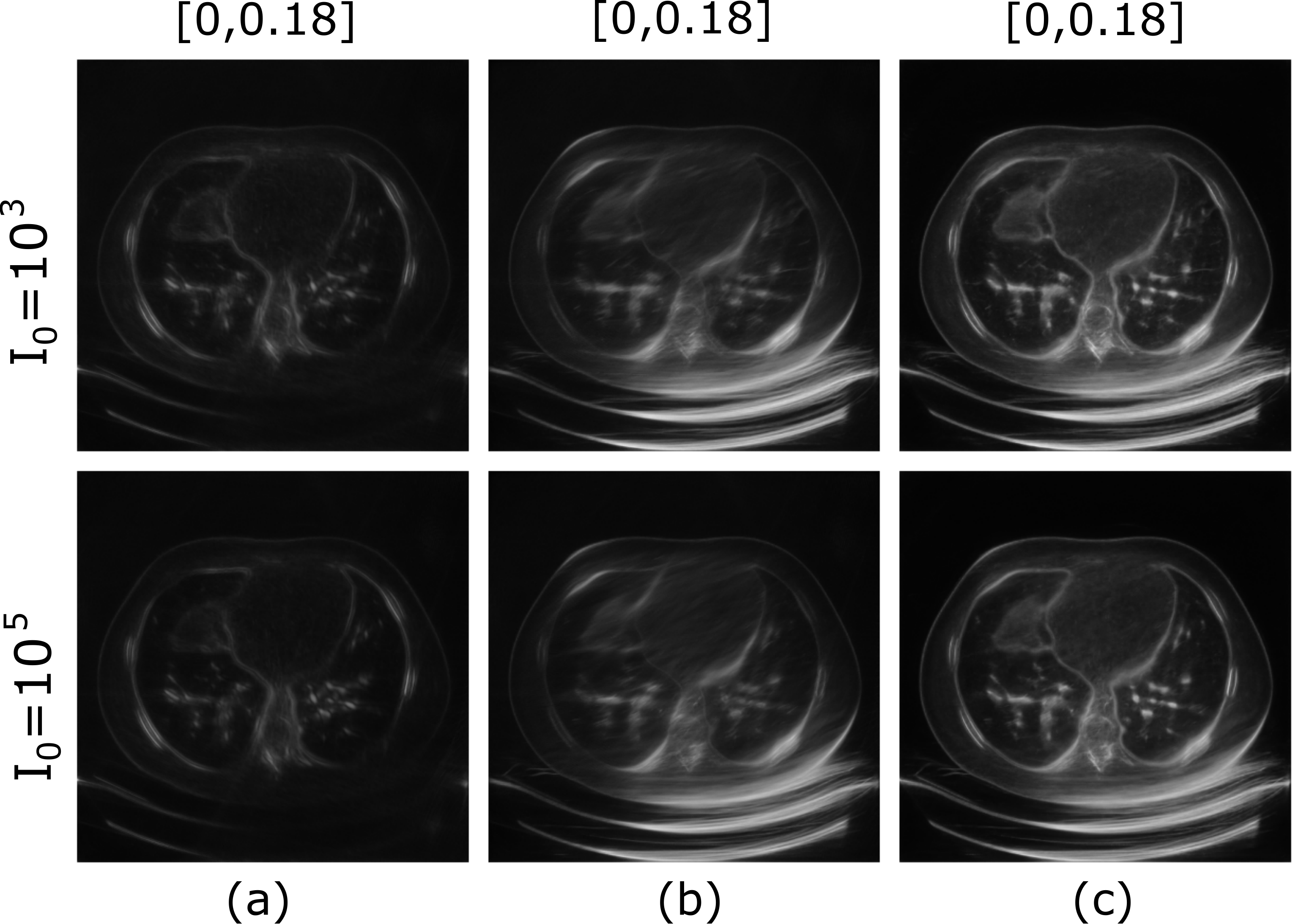}
     %  \vspace{-0.8cm}
      \caption{Uncertainty maps (a) $\cUMmeas$, (b) $\cUMnull$ and (c) $\cUM$ obtained with PULSE++ from CT measurements corresponding to Lung 1 for $I_0=10^3$ and $I_0=10^5$. The grayscale range of each type of uncertainty map is shown at the top of the corresponding column. It is evident that in both the cases, the uncertainty is primarily in the null space component. The variability in the measurable component is higher for $I_0=10^3$ due to a lower SNR in the projection data.}
      \label{fig:ct_uncertainty_maps}
\end{figure}

Uncertainty maps were computed for $T=100$ alternate solutions corresponding to projection data from Lung 1 and Lung 2 with $I_0=10^3$ and $I_0=10^5$ obtained using PULSE++, along with their measurable and null space components in the domain of the fan-beam projector $\vec{H}$. The uncertainty maps for Lung 1 are shown in Fig. \ref{fig:ct_uncertainty_maps}. Again, it was observed that the uncertainty in the alternate solutions was primarily due to variations in their null space component. The total uncertainty for the alternate solutions and their measurable and null space components for both Lung 1 and Lung 2 are shown in Table \ref{tab:ct_reconstruction_risk}. Since the projection data for $I_0=10^5$ had higher SNR, the uncertainty FOM in the measurable component was expectedly lower as compared to that for $I_0=10^3$. 

\begin{table}[!htbp]
    \centering
    \caption{Summary of uncertainty FOMs of alternate solutions obtained with PULSE++ from the same projection data using different values of $I_0$}
    \begin{tabular}{cc|ccc}
    \toprule
     Object & $I_0$ &$\| \cUMmeas \|^2_2$ &$\| \cUMnull \|^2_2$ &$\| \cUM \|^2_2$\\
     \midrule
     \multirow{2}{*}{Lung 1}
     &$10^3$   &63.72  &355.73  &420.04 \\
     &$10^5$   &49.80  &310.43  &360.74 \\
     \midrule
     
     \multirow{2}{*}{Lung 2}
     &$10^3$   &65.91  &468.35  &536.67 \\
     &$10^5$   &51.51  &408.71  &460.81 \\
     \bottomrule
    \end{tabular}
    \label{tab:ct_reconstruction_risk}
\end{table}
 
%\vspace{-0.3cm}
\section{Summary and conclusion}
\label{sec:discussion}
In this work, an empirical sampling method, called PULSE++, was proposed that employed generative model-constrained reconstruction with a StyleGAN to obtain multiple objects that are consistent with the same acquired tomographic measurement data. 
The proposed method represents an extension of the PULSE method that was originally developed for single image super-resolution applications, but employs improved statistical assumptions regarding the StyleGAN latent space.  It was demonstrated that the PULSE++ method was able to find data-consistent objects whereas the PULSE method could not. Additionally, it was illustrated that a state-of-the-art posterior sampling technique that employed a score-based generative model and annealed Langevin dynamics was unable to generate data-consistent alternate solutions. Uncertainty maps were computed, and it was observed that the PULSE++ method consistently estimated higher uncertainty in the null space component and  lower uncertainty in the measurable space component compared to the other methods.  Moreover, the SGMLS method significantly underestimated uncertainty, which further establishes the need for new efficient empirical sampling methods such as PULSE++ to perform reliable uncertainty quantification. 

The proposed PULSE++ method is general and, in principle, can be applied to any tomographic imaging system. While the studies in this paper concerned with two-dimensional imaging systems, the PULSE++ method can be extended to three-dimensional imaging systems by use of three-dimensional StyleGAN architectures \cite{hong20213d}. Furthermore, the proposed framework may be readily adapted for use with future style-based deep generative models \cite{karras2020analyzing,Karras2021}.

The use of a StyleGAN in the PULSE++ method presents certain challenges.
The StyleGAN must be sufficiently well-trained and should  accurately represent the to-be-imaged object distribution.  This can be challenging in diagnostic imaging applications, where the objects can contain varying pathologies that may not be fully represented in the StyleGAN training data \cite{skandarani2021gans,deshpande2021method}.  As such, the representation error of the StyleGAN should be acknowledged when employing PULSE++. However, because the PULSE++ method is intended to facilitate early stage assessments of new imaging technologies, this representation error may be more tolerable than it would be if it was intended as an approximate Bayesian reconstruction method for clinical use.

There remain additional topics for future studies. 
In the presented studies, the true imaging operator was assumed to be known. When the goal is to facilitate the  assessment of imaging technologies in virtual imaging studies, this may not be a limitation.
 However,  the impact of modeling errors on the performance of the PULSE++ method when applied to experimental measurements remains an important topic for investigation \cite{ravishankar2019image}. Additionally, it will be important to explore the application of the PULSE++ method for analyzing image reconstruction instabilities \cite{gottschling2020troublesome,bhadra2021hallucinations} and enabling adaptive imaging procedures \cite{barrett2008adaptive,clarkson2008task}.

\appendix
Additional alternate solutions for each of the numerical studies in Sec. \ref{sec:results} are available in the form of .mp4 video files under the ``Supplementary Files'' tab on ScholarOne\textregistered\ Manuscripts.

\bibliography{references}{}

% Generated by IEEEtran.bst, version: 1.14 (2015/08/26)
\begin{thebibliography}{10}
\providecommand{\url}[1]{#1}
\csname url@samestyle\endcsname
\providecommand{\newblock}{\relax}
\providecommand{\bibinfo}[2]{#2}
\providecommand{\BIBentrySTDinterwordspacing}{\spaceskip=0pt\relax}
\providecommand{\BIBentryALTinterwordstretchfactor}{4}
\providecommand{\BIBentryALTinterwordspacing}{\spaceskip=\fontdimen2\font plus
\BIBentryALTinterwordstretchfactor\fontdimen3\font minus
  \fontdimen4\font\relax}
\providecommand{\BIBforeignlanguage}[2]{{%
\expandafter\ifx\csname l@#1\endcsname\relax
\typeout{** WARNING: IEEEtran.bst: No hyphenation pattern has been}%
\typeout{** loaded for the language `#1'. Using the pattern for}%
\typeout{** the default language instead.}%
\else
\language=\csname l@#1\endcsname
\fi
#2}}
\providecommand{\BIBdecl}{\relax}
\BIBdecl

\bibitem{kak2002principles}
A.~C. Kak, M.~Slaney, and G.~Wang, ``Principles of computerized tomographic
  imaging,'' \emph{Medical Physics}, vol.~29, no.~1, pp. 107--107, 2002.

\bibitem{zbontar2018fastmri}
J.~Zbontar, F.~Knoll, A.~Sriram, M.~J. Muckley, M.~Bruno, A.~Defazio,
  M.~Parente, K.~J. Geras, J.~Katsnelson, H.~Chandarana \emph{et~al.},
  ``fast{MRI}: An open dataset and benchmarks for accelerated {MRI},''
  \emph{arXiv preprint arXiv:1811.08839}, 2018.

\bibitem{candes2006stable}
E.~J. Candes, J.~K. Romberg, and T.~Tao, ``Stable signal recovery from
  incomplete and inaccurate measurements,'' \emph{Communications on Pure and
  Applied Mathematics: A Journal Issued by the Courant Institute of
  Mathematical Sciences}, vol.~59, no.~8, pp. 1207--1223, 2006.

\bibitem{sidky2008image}
E.~Y. Sidky and X.~Pan, ``Image reconstruction in circular cone-beam computed
  tomography by constrained, total-variation minimization,'' \emph{Physics in
  Medicine \& Biology}, vol.~53, no.~17, p. 4777, 2008.

\bibitem{ravishankar2019image}
S.~Ravishankar, J.~C. Ye, and J.~A. Fessler, ``Image reconstruction: From
  sparsity to data-adaptive methods and machine learning,'' \emph{Proceedings
  of the IEEE}, vol. 108, no.~1, pp. 86--109, 2019.

\bibitem{kelly2017deep}
B.~Kelly, T.~P. Matthews, and M.~A. Anastasio, ``Deep learning-guided image
  reconstruction from incomplete data,'' \emph{arXiv preprint
  arXiv:1709.00584}, 2017.

\bibitem{mccann2017convolutional}
M.~T. McCann, K.~H. Jin, and M.~Unser, ``Convolutional neural networks for
  inverse problems in imaging: A review,'' \emph{IEEE Signal Processing
  Magazine}, vol.~34, no.~6, pp. 85--95, 2017.

\bibitem{bhadra2021hallucinations}
S.~Bhadra, V.~A. Kelkar, F.~J. Brooks, and M.~A. Anastasio, ``On hallucinations
  in tomographic image reconstruction,'' \emph{IEEE transactions on medical
  imaging}, vol.~40, no.~11, pp. 3249--3260, 2021.

\bibitem{tick2016image}
J.~Tick, A.~Pulkkinen, and T.~Tarvainen, ``Image reconstruction with
  uncertainty quantification in photoacoustic tomography,'' \emph{The Journal
  of the Acoustical Society of America}, vol. 139, no.~4, pp. 1951--1961, 2016.

\bibitem{shaw2021estimating}
R.~Shaw, C.~H. Sudre, S.~Ourselin, and M.~J. Cardoso, ``Estimating {MRI} image
  quality via image reconstruction uncertainty,'' \emph{arXiv preprint
  arXiv:2106.10992}, 2021.

\bibitem{gottschling2020troublesome}
N.~M. Gottschling, V.~Antun, B.~Adcock, and A.~C. Hansen, ``The troublesome
  kernel: why deep learning for inverse problems is typically unstable,''
  \emph{arXiv preprint arXiv:2001.01258}, 2020.

\bibitem{clarkson2008task}
E.~Clarkson, M.~A. Kupinski, H.~H. Barrett, and L.~Furenlid, ``A task-based
  approach to adaptive and multimodality imaging,'' \emph{Proceedings of the
  IEEE}, vol.~96, no.~3, pp. 500--511, 2008.

\bibitem{barrett2008adaptive}
H.~H. Barrett, L.~R. Furenlid, M.~Freed, J.~Y. Hesterman, M.~A. Kupinski,
  E.~Clarkson, and M.~K. Whitaker, ``Adaptive spect,'' \emph{IEEE transactions
  on medical imaging}, vol.~27, no.~6, pp. 775--788, 2008.

\bibitem{ulrych2001bayes}
T.~J. Ulrych, M.~D. Sacchi, and A.~Woodbury, ``A bayes tour of inversion: A
  tutorial,'' \emph{Geophysics}, vol.~66, no.~1, pp. 55--69, 2001.

\bibitem{mohebi2006posterior}
A.~Mohebi and P.~Fieguth, ``Posterior sampling of scientific images,'' in
  \emph{International Conference Image Analysis and Recognition}.\hskip 1em
  plus 0.5em minus 0.4em\relax Springer, 2006, pp. 339--350.

\bibitem{green2015bayesian}
P.~J. Green, K.~{\L}atuszy{\'n}ski, M.~Pereyra, and C.~P. Robert, ``{B}ayesian
  computation: a summary of the current state, and samples backwards and
  forwards,'' \emph{Statistics and Computing}, vol.~25, no.~4, pp. 835--862,
  2015.

\bibitem{sun2020deep}
H.~Sun and K.~L. Bouman, ``Deep probabilistic imaging: Uncertainty
  quantification and multi-modal solution characterization for computational
  imaging,'' \emph{arXiv preprint arXiv:2010.14462}, vol.~9, 2020.

\bibitem{herrmann2021KAUSTdbi}
A.~Siahkoohi, G.~Rizzuti, M.~Louboutin, P.~A. Witte, and F.~J. Herrmann, ``Deep
  {B}ayesian inference for task-based seismic imaging,'' in \emph{KAUST}, 03
  2021, talk at KAUST.

\bibitem{mosser2018stochastic}
L.~Mosser, O.~Dubrule, and M.~J. Blunt, ``Stochastic seismic waveform inversion
  using generative adversarial networks as a geological prior,'' \emph{arXiv
  preprint arXiv:1806.03720}, 2018.

\bibitem{jalal2021robust}
A.~Jalal, M.~Arvinte, G.~Daras, E.~Price, A.~G. Dimakis, and J.~I. Tamir,
  ``Robust compressed sensing {MRI} with deep generative priors,'' \emph{arXiv
  preprint arXiv:2108.01368}, 2021.

\bibitem{bardsley2014randomize}
J.~M. Bardsley, A.~Solonen, H.~Haario, and M.~Laine, ``Randomize-then-optimize:
  A method for sampling from posterior distributions in nonlinear inverse
  problems,'' \emph{SIAM Journal on Scientific Computing}, vol.~36, no.~4, pp.
  A1895--A1910, 2014.

\bibitem{akiyama2019first}
K.~Akiyama, A.~Alberdi, W.~Alef, K.~Asada, R.~Azulay, A.-K. Baczko, D.~Ball,
  M.~Balokovi{\'c}, J.~Barrett, D.~Bintley \emph{et~al.}, ``First m87 event
  horizon telescope results. iv. imaging the central supermassive black hole,''
  \emph{The Astrophysical Journal Letters}, vol. 875, no.~1, p.~L4, 2019.

\bibitem{spall2012stochastic}
J.~C. Spall, ``Stochastic optimization,'' in \emph{Handbook of computational
  statistics}.\hskip 1em plus 0.5em minus 0.4em\relax Springer, 2012, pp.
  173--201.

\bibitem{goodfellow2016deep}
I.~Goodfellow, Y.~Bengio, A.~Courville, and Y.~Bengio, \emph{Deep
  learning}.\hskip 1em plus 0.5em minus 0.4em\relax MIT press Cambridge, 2016,
  vol.~1.

\bibitem{menon2020pulse}
S.~Menon, A.~Damian, S.~Hu, N.~Ravi, and C.~Rudin, ``Pulse: Self-supervised
  photo upsampling via latent space exploration of generative models,'' in
  \emph{Proceedings of the ieee/cvf conference on computer vision and pattern
  recognition}, 2020, pp. 2437--2445.

\bibitem{karras2019style}
T.~Karras, S.~Laine, and T.~Aila, ``A style-based generator architecture for
  generative adversarial networks,'' in \emph{Proceedings of the IEEE
  Conference on Computer Vision and Pattern Recognition}, 2019, pp. 4401--4410.

\bibitem{barrett2013foundations}
H.~H. Barrett and K.~J. Myers, \emph{Foundations of {I}mage {S}cience}.\hskip
  1em plus 0.5em minus 0.4em\relax John Wiley \& Sons, 2013.

\bibitem{zhou2021learning}
W.~Zhou, S.~Bhadra, F.~J. Brooks, H.~Li, and M.~A. Anastasio, ``Learning
  stochastic object models from medical imaging measurements by use of advanced
  {A}mbient{GAN}s,'' \emph{arXiv preprint arXiv:2106.14324}, 2021.

\bibitem{hong20213d}
S.~Hong, R.~Marinescu, A.~V. Dalca, A.~K. Bonkhoff, M.~Bretzner, N.~S. Rost,
  and P.~Golland, ``3d-{S}tyle{GAN}: A style-based generative adversarial
  network for generative modeling of three-dimensional medical images,'' in
  \emph{Deep Generative Models, and Data Augmentation, Labelling, and
  Imperfections}.\hskip 1em plus 0.5em minus 0.4em\relax Springer, 2021, pp.
  24--34.

\bibitem{huang2017arbitrary}
X.~Huang and S.~Belongie, ``Arbitrary style transfer in real-time with adaptive
  instance normalization,'' in \emph{Proceedings of the IEEE International
  Conference on Computer Vision}, 2017, pp. 1501--1510.

\bibitem{kelkar2021prior}
V.~A. Kelkar and M.~A. Anastasio, ``Prior image-constrained reconstruction
  using style-based generative models,'' \emph{arXiv preprint
  arXiv:2102.12525}, 2021.

\bibitem{fetty2020latent}
L.~Fetty, M.~Bylund, P.~Kuess, G.~Heilemann, T.~Nyholm, D.~Georg, and
  T.~L{\"o}fstedt, ``Latent space manipulation for high-resolution medical
  image synthesis via the {S}tyle{GAN},'' \emph{Zeitschrift f{\"u}r
  Medizinische Physik}, vol.~30, no.~4, pp. 305--314, 2020.

\bibitem{schutte2021using}
K.~Schutte, O.~Moindrot, P.~H{\'e}rent, J.-B. Schiratti, and S.~J{\'e}gou,
  ``Using {S}tyle{GAN} for visual interpretability of deep learning models on
  medical images,'' \emph{arXiv preprint arXiv:2101.07563}, 2021.

\bibitem{abdal2019image2stylegan}
R.~Abdal, Y.~Qin, and P.~Wonka, ``Image2{S}tylegan: How to embed images into
  the {S}tyle{GAN} latent space?'' in \emph{Proceedings of the IEEE/CVF
  International Conference on Computer Vision}, 2019, pp. 4432--4441.

\bibitem{wulff2020improving}
J.~Wulff and A.~Torralba, ``Improving inversion and generation diversity in
  {S}tyle{GAN} using a {G}aussianized latent space,'' \emph{arXiv preprint
  arXiv:2009.06529}, 2020.

\bibitem{zhu2020improved}
P.~Zhu, R.~Abdal, Y.~Qin, J.~Femiani, and P.~Wonka, ``Improved style{GAN}
  embedding: Where are the good latents?'' \emph{arXiv preprint
  arXiv:2012.09036}, 2020.

\bibitem{bora2017compressed}
A.~Bora, A.~Jalal, E.~Price, and A.~G. Dimakis, ``Compressed sensing using
  generative models,'' in \emph{Proceedings of the 34th International
  Conference on Machine Learning-Volume 70}.\hskip 1em plus 0.5em minus
  0.4em\relax JMLR. org, 2017, pp. 537--546.

\bibitem{bhadra2020medical}
S.~Bhadra, W.~Zhou, and M.~A. Anastasio, ``Medical image reconstruction with
  image-adaptive priors learned by use of generative adversarial networks,'' in
  \emph{Medical Imaging 2020: Physics of Medical Imaging}, vol. 11312.\hskip
  1em plus 0.5em minus 0.4em\relax International Society for Optics and
  Photonics, 2020, p. 113120V.

\bibitem{kelkar2021compressible}
V.~A. Kelkar, S.~Bhadra, and M.~A. Anastasio, ``Compressible latent-space
  invertible networks for generative model-constrained image reconstruction,''
  \emph{IEEE Transactions on Computational Imaging}, 2021.

\bibitem{kingma2014adam}
D.~P. Kingma and J.~Ba, ``Adam: {A} method for stochastic optimization,''
  \emph{arXiv preprint arXiv:1412.6980}, 2014.

\bibitem{stylegan}
\BIBentryALTinterwordspacing
T.~Karras, S.~Laine, and T.~Aila, ``Style{GAN} - official {T}ensor{F}low
  implementation,'' 2018. [Online]. Available:
  \url{https://github.com/NVlabs/stylegan}
\BIBentrySTDinterwordspacing

\bibitem{yan2018deeplesion}
K.~Yan, X.~Wang, L.~Lu, and R.~M. Summers, ``Deeplesion: automated mining of
  large-scale lesion annotations and universal lesion detection with deep
  learning,'' \emph{Journal of Medical Imaging}, vol.~5, no.~3, p. 036501,
  2018.

\bibitem{pulse_pp}
\BIBentryALTinterwordspacing
S.~Bhadra, U.~Villa, and M.~A. Anastasio, ``Mining the manifolds of deep
  generative models for multiple data-consistent solutions of tomographic
  imaging problems - py{T}orch implementation,'' 2022. [Online]. Available:
  \url{https://github.com/comp-imaging-sci/mining-tomo-solutions-pulse}
\BIBentrySTDinterwordspacing

\bibitem{heusel2017gans}
M.~Heusel, H.~Ramsauer, T.~Unterthiner, B.~Nessler, and S.~Hochreiter, ``Gans
  trained by a two time-scale update rule converge to a local nash
  equilibrium,'' in \emph{Advances in Neural Information Processing Systems},
  2017, pp. 6626--6637.

\bibitem{skandarani2021gans}
Y.~Skandarani, P.-M. Jodoin, and A.~Lalande, ``{GAN}s for medical image
  synthesis: An empirical study,'' \emph{arXiv preprint arXiv:2105.05318},
  2021.

\bibitem{deshpande2021method}
R.~Deshpande, M.~A. Anastasio, and F.~J. Brooks, ``A method for evaluating the
  capacity of generative adversarial networks to reproduce high-order spatial
  context,'' \emph{arXiv preprint arXiv:2111.12577}, 2021.

\bibitem{borji2021pros}
A.~Borji, ``Pros and cons of {GAN} evaluation measures: New developments,''
  \emph{arXiv preprint arXiv:2103.09396}, 2021.

\bibitem{kelkar2022assessing}
V.~A. Kelkar, D.~S. Gotsis, F.~J. Brooks, P.~KC, K.~J. Myers, R.~Zeng, and
  M.~A. Anastasio, ``Assessing the ability of generative adversarial networks
  to learn canonical medical image statistics,'' \emph{arXiv preprint
  arXiv:2204.12007}, 2022.

\bibitem{vershynin2018random}
R.~Vershynin, ``Random vectors in high dimensions,'' \emph{Cambridge Series in
  Statistical and Probabilistic Mathematics. Cambridge University Press},
  vol.~3, pp. 38--69, 2018.

\bibitem{reddi2019convergence}
S.~J. Reddi, S.~Kale, and S.~Kumar, ``On the convergence of {A}dam and
  beyond,'' \emph{arXiv preprint arXiv:1904.09237}, 2019.

\bibitem{pulse}
\BIBentryALTinterwordspacing
S.~Menon, A.~Damian, S.~Hu, N.~Ravi, and C.~Rudin, ``Pulse: Self-supervised
  photo upsampling via latent space exploration of generative models,'' 2020.
  [Online]. Available: \url{https://github.com/adamian98/pulse}
\BIBentrySTDinterwordspacing

\bibitem{tf2torch}
\BIBentryALTinterwordspacing
P.~Bialecki and T.~Viehmann, ``Py{T}orch implementation of the {S}tyle{GAN}
  generator,'' 2019. [Online]. Available:
  \url{https://github.com/lernapparat/lernapparat/tree/master/style_gan}
\BIBentrySTDinterwordspacing

\bibitem{song2019generative}
Y.~Song and S.~Ermon, ``Generative modeling by estimating gradients of the data
  distribution,'' \emph{Advances in Neural Information Processing Systems},
  vol.~32, 2019.

\bibitem{song2020improved}
------, ``Improved techniques for training score-based generative models,''
  \emph{Advances in neural information processing systems}, vol.~33, pp.
  12\,438--12\,448, 2020.

\bibitem{langevin}
\BIBentryALTinterwordspacing
A.~Jalal, M.~Arvinte, G.~Daras, E.~Price, A.~G. Dimakis, and J.~I. Tamir,
  ``{CSGM}-{MRI}-{L}angevin,'' 2021. [Online]. Available:
  \url{https://github.com/utcsilab/csgm-mri-langevin}
\BIBentrySTDinterwordspacing

\bibitem{aja2016statistical}
S.~Aja-Fern{\'a}ndez and G.~Vegas-S{\'a}nchez-Ferrero, ``Statistical analysis
  of noise in {MRI},'' \emph{Switzerland: Springer International Publishing},
  2016.

\bibitem{fastmri}
\BIBentryALTinterwordspacing
J.~e.~a. Zbontar, ``fast{MRI},'' 2018. [Online]. Available:
  \url{https://github.com/facebookresearch/fast{MRI}}
\BIBentrySTDinterwordspacing

\bibitem{morozov1966solution}
V.~A. Morozov, ``On the solution of functional equations by the method of
  regularization,'' in \emph{Doklady Akademii Nauk}, vol. 167, no.~3.\hskip 1em
  plus 0.5em minus 0.4em\relax Russian Academy of Sciences, 1966, pp. 510--512.

\bibitem{ma2022accelerating}
H.~Ma, L.~Zhang, X.~Zhu, J.~Zhang, and J.~Feng, ``Accelerating score-based
  generative models for high-resolution image synthesis,'' \emph{arXiv preprint
  arXiv:2206.04029}, 2022.

\bibitem{ding2016modeling}
Q.~Ding, Y.~Long, X.~Zhang, and J.~A. Fessler, ``Modeling mixed
  {P}oisson-{G}aussian noise in statistical image reconstruction for {X}-ray
  {CT},'' \emph{Arbor}, vol. 1001, p. 48109, 2016.

\bibitem{hansen2018air}
P.~C. Hansen and J.~S. J{\o}rgensen, ``Air tools ii: algebraic iterative
  reconstruction methods, improved implementation,'' \emph{Numerical
  Algorithms}, vol.~79, no.~1, pp. 107--137, 2018.

\bibitem{leng2019photon}
S.~Leng, M.~Bruesewitz, S.~Tao, K.~Rajendran, A.~F. Halaweish, N.~G. Campeau,
  J.~G. Fletcher, and C.~H. McCollough, ``Photon-counting detector {CT}: system
  design and clinical applications of an emerging technology,''
  \emph{Radiographics}, vol.~39, no.~3, pp. 729--743, 2019.

\bibitem{sixou2018morozov}
B.~Sixou, T.~Hohweiller, and N.~Ducros, ``Morozov principle for
  {K}ullback-{L}eibler residual term and {P}oisson noise,'' \emph{Inverse
  Problems \& Imaging}, vol.~12, no.~3, p. 607, 2018.

\bibitem{karras2020analyzing}
T.~Karras, S.~Laine, M.~Aittala, J.~Hellsten, J.~Lehtinen, and T.~Aila,
  ``Analyzing and improving the image quality of {S}tyle{GAN},'' in
  \emph{Proceedings of the IEEE/CVF Conference on Computer Vision and Pattern
  Recognition}, 2020, pp. 8110--8119.

\bibitem{Karras2021}
T.~Karras, M.~Aittala, S.~Laine, E.~H\"ark\"onen, J.~Hellsten, J.~Lehtinen, and
  T.~Aila, ``Alias-free generative adversarial networks,'' in \emph{Proc.
  NeurIPS}, 2021.

\end{thebibliography}
\bibliographystyle{IEEEtran}

\end{document}